%% file: multiharmonic.tex
\documentclass[useAMS,usenatbib,usegraphicx]{mn2e}
\usepackage{amsmath, hyperref, amssymb, color, graphicx, latexsym}
\usepackage{multirow}
\usepackage{xspace}
\usepackage{algorithm, algpseudocode, varwidth}
\usepackage{subcaption}

\DeclareGraphicsExtensions{%
    .pdf,%
    .png,%
    .jpg,.mps,.jpeg,.jbig2,.jb2,.JPG,.JPEG,.JBIG2,.JB2}

\newcommand{\gws}{gravitational waves\xspace}
\newcommand{\gw}{gravitational wave\xspace}
\newcommand{\snr}{SNR\xspace}
\newcommand{\snrs}{SNRs\xspace}
\newcommand{\ta}{triaxial aligned\xspace}
\newcommand{\tn}{triaxial non-aligned\xspace}
\newcommand{\biax}{biaxial\xspace}

\newcommand{\roughly}{\mathchar"5218\relax} 
\newcommand{\rmi}{\mathrm{i}}
\newcommand{\rmd}{\mathrm{d}}
\newcommand{\rme}{\mathrm{e}}
\newcommand{\const}{\mathrm{const.}}

\title[Dual-harmonic GWs from NSs]{First results and future prospects for dual-harmonic
searches for gravitational waves from spinning neutron stars}
\author[M.\ Pitkin et al]{
M.~Pitkin,$^1$\thanks{matthew.pitkin@glasgow.ac.uk}
C.~Gill,$^1$\thanks{Now at Selex}
D.~I.~Jones,$^2$\thanks{d.i.jones@soton.ac.uk}
G.~Woan,$^1$\thanks{graham.woan@glasgow.ac.uk}
and G.~S.~Davies$^1$\thanks{g.davies.2@research.gla.ac.uk} \\
$^1$SUPA, School of Physics and Astronomy, University of Glasgow, University Avenue, Glasgow, G12
8QQ, UK \\
$^2$Mathematical Sciences and STAG Research Centre, University of Southampton, Southampton, SO17
1BJ, UK}

\date{}

\pagerange{\pageref{firstpage}--\pageref{lastpage}} \pubyear{2015}

\begin{document}

\maketitle

\begin{abstract}
\input{abstract.tex}
\end{abstract}


\section{Introduction}

Several searches have been performed for \gws from known pulsars in data from the LIGO, GEO600 and
Virgo \gw detectors
\citep{Abbott:2005,Abbott:2007,Abbott:2008,Abbott:2010,Abadie:2011,2014ApJ...785..119A}.
These rely on the known phase evolution of the pulsars from electromagnetic observations
\citep[e.g.][]{Manchester:2005} to allow long duration (of order a year) coherent searches for
signals from them in \gw data. Unfortunately no signal has yet been seen, but interesting upper
limits on \gw amplitude have been produced, and for two pulsars (the Crab and Vela pulsars) the
``spin-down limit'' has been beaten \citep{Abbott:2008,Abadie:2011}. One of the principal previous
methods used for these searches \citep{Dupuis:2005} has focussed on parameter estimation and the
setting of upper limits, but has not provided any measure of detection confidence.

Previous \gw searches targeting known pulsars have assumed \gw emission at a single frequency,
taken to be twice (or very close to twice) the spin frequency. However, there are reasons to
consider slightly more general
waveforms. In this paper we consider the model proposed by \citet{Jones:2010}, dealing with
steadily rotating triaxial stars. Specifically, \citet{Jones:2010} considered a star containing a
pinned superfluid. Such a star can rotate steadily about an axis that does not coincide with the
principal axis of the solid crust, and will generically emit gravitational radiation at both the
spin frequency $f$ and at $2f$. We term this the \emph{\tn} case. This contrasts with the
`standard' scenario, considered in almost all \gw searches to date, of rotation about a principal
axis, which emits only at $2f$. We term this the \emph{\ta} case, and it can be regarded as a
special case of the \tn case. Another special case is that of a \emph{\biax} star, where two moments of
inertia of the star are equal. The waveform in this case is identical to that of a biaxial
precessing star of the sort considered by \citet{Zimmermann:1979}, which also produces
gravitational waves at two frequencies. However, precession generically results in a modulation in
the electromagnetic signal produced by a pulsar \citep[see e.g.][]{JA:2001}, something that is not
clearly observed in the pulsar population. In contrast, in the model of \citet{Jones:2010}, there
is emission at $f$ and $2f$ even in a steadily spinning star. The attraction of this model is that
such emission, at both $f$ and $2f$, \emph{might} be being produced by any of the known pulsars,
without leaving any tell-tale signature in the radio pulsations. It is therefore clearly of
interest to understand the issues that arise when carrying out gravitational wave searches for such
double-component signals.

In this paper we study how different parameterisations of the model affect the estimation of
signal parameters and the astrophysical information that can be extracted. We also discuss applying
Bayesian model selection to assess the detection of signals from these sources and perform
comparisons between the different signal models.
A similar study has been performed by \citet{Bejger:2014} although there analysis was based on a
maximum likelihood approach to parameter estimation.  We use the methods we have developed to analyse data from LIGO's fifth science run (S5), setting upper limits on the emission at both $f$ and $2f$ for 43 known pulsars.

The plan of this paper is as follows. In Section~\ref{sect:model} we give a brief description of
the neutron star model and waveform, confining the details to Appendix~\ref{sect:relating_params}.
In Section~\ref{sect:bayesian} we describe the Bayesian methodology we employ. In
Section~\ref{sect:waveform_v_source} we briefly look at the shape of the parameter probability
distributions for two different signal parameterisations. In Section~\ref{sect:model_selection}
we show how these Bayesian methods can allow us to distinguish between the three different sorts of
signals described above. In Section~\ref{sec:S5results} we present the results from applying a
search for \gw emission at both $f$ and $2f$ in LIGO data. We summarise our findings in Section~\ref{sect:conclusions}.

\section{The model} \label{sect:model}

In this Section we describe the physical model and gravitational wave emission from our triaxial
star. In Section~\ref{sect:signal_source_params} we use the original parameterisation of \citet{Jones:2010},
while in Section~\ref{sect:signal_waveform_params} we use an alternative simpler set of parameters,
as described in \citet{Jones:2015}. The ranges of the relevant parameters are given in
Section~\ref{sect:param_ranges}, again based on the analysis of \citet{Jones:2015}.

\subsection{The signal written in terms of source parameters}
\label{sect:signal_source_params}

Here we recap the physical model given in \citet{Jones:2010}.  The neutron star is triaxial, with a
moment of inertia tensor whose principal components are $(I_1, I_2, I_3)$.  Because of superfluid
pinning, it can rotate about an axis, fixed in the inertial frame, that does not coincide with any
one of these principal axes. This gives rise to gravitational wave emission at both $f$ and $2f$. The signal
in a detector at the rotation frequency ($f$) is \citep{Jones:2010, Jones:2015}
\begin{align}\label{eq:1f}
h^{f}(t) =&
F_{+}(\psi,t)\sin{\iota}\cos{\iota}\Big\{I_{21}\sin{2\lambda}\sin{\theta}\cos{
\phi(t)}
+ \nonumber \\
 & (I_{21}\cos{}^2{\lambda} - I_{31})\sin{2\theta}\sin{\phi(t)}\Big\}
\nonumber \\
 & - F_{\times}(\psi,t)\sin{\iota}\Big\{(I_{21}\cos{}^2{\lambda} -
I_{31})\sin{2\theta}\cos{\phi(t)} -
\nonumber \\
 & I_{21}\sin{2\lambda}\sin{\theta}\sin{\phi(t)}\Big\},
\end{align}
and the signal at twice the rotation frequency ($2f$) is
\begin{align}\label{eq:2f}
h^{2f}(t) =& 2F_{+}(\psi,t)(1+\cos{}^2\iota)\Big\{\big[I_{21}(\sin{}^2\lambda -
\cos{}^2\lambda
\cos{}^2\theta) -
\nonumber \\
& I_{31}\sin{}^2\theta\big]\cos{2\phi(t)} +
I_{21}\sin{2\lambda}\cos{\theta}\sin{2\phi(t)}\Big\}
\nonumber \\
 &
- 4F_{\times}(\psi,t)\cos{\iota}\Big\{I_{21}\sin{2\lambda}\cos{\theta}\cos{
2\phi(t)} - \nonumber \\
 & \big[I_{21}(\sin{}^2\lambda - \cos{}^2\lambda\cos{}^2\theta) -
I_{31}\sin{}^2\theta\big]\sin{2\phi(t)}\Big\}.
\end{align}
The polarisation factors $F_+$ and $F_\times$ depend upon the polarisation angle $\psi$ of the
source. They also depend on the position of the source on the sky.
We have not explicitly labelled this dependence as these parameters would be known for a
targeted gravitational wave search. The angle $\iota$ is the inclination angle of the star's spin
vector with respect to the observer.

The evolution in phase $\phi(t)$ is generated by the rotation
of the star, so that $\phi(t) = 2\pi \int_{t_0}^t f(t') \, \rmd t' + \phi_0$ where $f(t)$ is the
frequency evolution and $\phi_0$ the phase at $t_0$.  In practice, for targeted gravitational wave searches,
 $f(t)$ will be a known function (known e.g.\ from radio pulsar observations), and so we will simply write
\begin{equation}
\phi(t) = \Omega t + \phi_0 ,
\end{equation}
treating $\Omega = 2\pi f$ as a constant.

The constant angles $(\theta, \phi_0, \lambda)$
are the Euler angles that specify the orientation of the star with respect to the
inertial frame (at time $t_0$).  Here we have used $\lambda$ to replace the `$\psi$' parameter in
\citet{Jones:2015} to avoid confusion with the standard use of $\psi$ for \gw polarisation angle.
The parameters $I_{21}$ and $I_{31}$ are measures of the asymmetry in the moment of inertia tensor,
with factors of the angular spin frequency $\Omega$ and distance $r$ absorbed for convenience:
\begin{equation}
I_{21} \equiv \frac{\Omega^2 (I_2-I_1)}{r} , \hspace{10mm} I_{31} \equiv
\frac{\Omega^2 (I_3-I_1)}{r} .
\end{equation}
Putting all of this together, and assuming that the sky position and spin frequency are
already known, we have a set of seven \emph{source parameters}:
\begin{equation}
\label{eq:source_params}
\btheta^a_{\rm source} = \{\iota, \psi, I_{21}, I_{31}, \theta, \phi_0, \lambda \}.
\end{equation}

We term this general case the \emph{\tn model} of a spinning neutron star. There are two special cases that we will single out.
The first is a triaxial star spinning about a principal axis. This can be obtained from Eqns~
(\ref{eq:1f}) and (\ref{eq:2f}) by setting $\theta=0$; there is then emission only at $2f$. We term
this the \emph{\ta} case. The second special case is the \emph{\biax} case, where two of the
principal components of the quadrupole moment tensor are equal. This can be found by setting
$I_{21}=0$, and produces emission at both $f$ and $2f$. Note that, physically, this is slightly
different from the relatively well-known \emph{precessional} motion of a biaxial star
\citep[see e.g.][]{Zimmermann:1979,Jones:2002}, as the latter has an additional slow rotation,
superimposed
about the symmetry axis.  However, the time variation of the mass quadrupole, and therefore the
corresponding gravitational  waveforms, are  identical in the two cases, so all of the discussion
of the biaxial case in this paper applies also to the biaxial precession waveform.  Nevertheless,
it should be remembered that in the precession case, there can be modulation in the observed
electromagnetic pulsation frequency, and the time average of this electromagnetic  pulsation
frequency can be offset from the gravitational wave frequency; see \citet{Jones:2002} for a detailed
discussion.

\subsection{The signal written in terms of waveform parameters}
\label{sect:signal_waveform_params}

As previously shown by one of us \citep{Jones:2015}, and also explained in \citet{Bejger:2014}, the
physical source model, specified by the seven parameters of Eqn.~(\ref{eq:source_params}),
contains a degeneracy. If we instead express the model as complex harmonic amplitudes we find that
Eqns.~(\ref{eq:1f}) and (\ref{eq:2f}) can be rewritten as
\begin{align}
h^{f}(t) =& -\frac{1}{2}F_{+}(\psi,
t)C_{21}\sin{\iota}\cos{\iota}\cos{\left(\phi(t)+\Phi_{21}^C\right)} - \nonumber
\\
 \label{eq:1f_waveform} & \frac{1}{2}F_{\times}(\psi,
t)C_{21}\sin{\iota}\sin{\left(\phi(t)+\Phi_{21}^C\right)},
\end{align}
and
\begin{align}
h^{2f}(t) =& -F_{+}(\psi, t)C_{22}[1+\cos{}^2\iota]\cos{\left(2\phi(t)+\Phi_{22}^C\right)} -
\nonumber \\
\label{eq:2f_waveform}  & 2F_{\times}(\psi,
t)C_{22}\cos{\iota}\sin{\left(2\phi(t)+\Phi_{22}^C\right)}.
\end{align}
There now appear two amplitude-like parameters $C_{21}$ and $C_{22}$ with corresponding phase
parameters $\Phi^{\rm C}_{21}$ and $\Phi^{\rm C}_{22}$.  Assuming that the sky location and spin
frequency are known, we can identify a set of $6$ \emph{waveform parameters}, one fewer than in the
case of the source parameters:
\begin{equation}
\label{eq:waveform_params}
\btheta^a_{\rm waveform} = \{\iota, \psi,  C_{21}, C_{22}, \Phi^{\rm C}_{21}, \Phi^{\rm C}_{22} \} .
\end{equation}
When expressed in terms of these waveform parameters, a problematic degeneracy is removed, as we
will illustrate in Section~\ref{sect:waveform_v_source} below. Comparing with the source parameters
of Eqn.~(\ref{eq:source_params}), we see that the two angles $(\iota, \psi)$ giving the
orientation of the star's spin axis relative to the observer are common to both sets. There is in
fact a rather complicated algebraic relationship between the five remaining source parameters
$(I_{21}, I_{31}, \theta, \phi_0, \lambda)$ and the four remaining waveform parameters $(C_{21},
C_{22}, \Phi^{\rm C}_{21}, \Phi^{\rm C}_{22})$. This relation is derived in \citet{Jones:2015}, and
reproduced in Appendix~\ref{sect:relating_params}, where we also summarise the form that the
waveform parameterisation takes when specialised to the \ta and \biax cases.   As shown in \cite{Jones:2015}, the parameters $C_{21}$ and $C_{22}$ are basically the (moduli) of the (complex) mass quadrupole scalars that describe the quadrupolar component of the mass distribution of the rotating star, with a factor of order $\Omega^2 /r$ absorbed for simplicity.  (Note that in \cite{Jones:2015} these quantities are denoted by $\tilde C_{21}, \tilde C_{22}$).


Note that in this analysis we are assuming a search for \gw signals from \emph{known} pulsars, or
sources where a significant \gw signal has already been found.  This means that, rather than
using the waveforms as written in Eqns~(\ref{eq:1f}--\ref{eq:2f}), or Eqns~(\ref{eq:1f_waveform}--\ref{eq:2f_waveform}),
we can remove the oscillations that take place at the
relatively high frequencies $f$ and $2f$, using the heterodyne method of \citet{Dupuis:2005} to give
instead  a pair of narrow-band complex times series. We do this by using the known phase
evolution of the signal $\phi(t)$, multiplying by $\rme^{-i\phi(t)}$ for the $f$-band and
$\rme^{-\rmi 2\phi(t)}$
for the $2f$-band. This heterodyning, and subsequent low-pass filtering, leaves a signal
model for the $f$ and $2f$ streams of
\begin{equation}\label{eq:hetf}
h_{f}(t) = -\frac{C_{21}}{4}F_+(\psi,t)\sin{\iota}\cos{\iota}\rme^{\rmi\Phi_{21}^C}
+
i\frac{C_{21}}{4}F_{\times}(\psi,t)\sin{\iota}\rme^{\rmi\Phi_{21}^C}
\end{equation}
and
\begin{equation}\label{eq:het2f}
h_{2f}(t) = -\frac{C_{22}}{2}F_+(\psi,t)[1+\cos{}^2\iota]\rme^{\rmi\Phi_{22}^C} +
iC_{22}F_{\times}(\psi,t)\cos{\iota}\rme^{\rmi\Phi_{22}^C}
\end{equation}
when written in terms of the waveform parameters.  A similar heterodyning can be applied to the
waveform when written in terms of the source parameters \citep[see][]{Gill:2012}.

\subsection{Parameter ranges} \label{sect:param_ranges}

In order to carry out our analyses, we need to choose sensible ranges in both the source and
waveform parameters, for each of the \ta, \biax and \tn models. The choice of ranges in these
parameters turns out to be rather subtle, and is described in detail in \citet{Jones:2015}.
Basically, the source parameterisation, and, to a lesser extent, the waveform parameterisation,
contain various \emph{discrete} degeneracies, where changes in some combination of angle and/or
amplitude parameters leaves the detected waveform $h(t)$ invariant. This allows the ranges in these
parameters to be reduced as compared to one's initial expectations,
with there being
several
options as to how the parameter space is reduced. The choices given in
Tables~\ref{tab:priorswaveform} and \ref{tab:priorssource} represent one of several possibilities
\citep[see][]{Jones:2015}, and are the ranges we have used for the subsequent analyses
presented in this paper. Any point in these ranges can be mapped into another part of the full
parameter range that gives an identical waveform through the transformations given in
\citet{Jones:2015}. This enables signal parameter estimation and evidence evaluation to be
performed using this minimal range, but for posteriors to then be mapped into the full
range, if so desired.

\begin{table}
\centering
\caption{Parameter ranges for the waveform parameters for the three signal models.}
\label{tab:priorswaveform}
\begin{tabular}{@{}lccc}
  \hline
  ~ & \multicolumn{3}{c}{Models} \\
  \hline
  ~ & Triaxial aligned & Biaxial & Triaxial non-aligned \\
  \hline
  $C_{21}$ & ---  & $-C_{21}^{\rm max}, C_{21}^{\rm max}$ & $0, C_{21}^{\rm max}$ \\
  $C_{22}$ & $0, C_{22}^{\rm max}$ & $-C_{22}^{\rm max}, C_{22}^{\rm max}$ & $0,
C_{22}^{\rm max}$ \\
  $\Phi_{21}^C$ (rads) & --- & $0, 2\pi$ & $0, 2\pi$ \\
  $\Phi_{22}^C$ (rads) & $0,2\pi$ & $2\Phi_{21}^C$ & $0, 2\pi$
\\
  $\psi$ (rads) & $0, \pi/2$ & $0, \pi/2$ & $0, \pi/2$ \\
  $\cos{\iota}$ & $-1, 1$ & $-1, 1$ & $-1, 1$ \\
  \hline
 \end{tabular}
\end{table}

\begin{table}
\centering
\caption{Parameter ranges for the source parameters for the three signal models.}
\label{tab:priorssource}
\begin{tabular}{@{}lccc}
  \hline
  ~ & \multicolumn{3}{c}{Models} \\
  \hline
  ~ & Triaxial aligned & Biaxial & Triaxial non-aligned \\
  \hline
  $I_{31}$ & --- & $0, I_{31}^{\rm max}$ & $I_{21}, I_{31}^{\rm max}$ \\
  $I_{21}$ & $0, I_{21}^{\rm max}$ & --- & $0, I_{21}^{\rm max}$ \\
  $\phi_{0}$ (rads) & $0, \pi$ & $0, 2\pi$ & $0, 2\pi$ \\
  $\lambda$ (rads) & --- & ---  & $0, \pi$ \\
  $\cos{\theta}$ & 1 & $0, 1$ & $0, 1$ \\
  $\psi$ (rads) & $0, \pi/2$ & $0, \pi$ & $0, \pi/2$ \\
  $\cos{\iota}$ & $-1, 1$ & $-1, 1$ & $-1, 1$ \\
  \hline
 \end{tabular}
\end{table}

\section{Bayesian methodology} \label{sect:bayesian}

For this analysis we want to be able to compute probability distributions for source and waveform
parameters, and also to compare models (noise-only verses triaxial non-aligned verses biaxial verses
triaxial aligned).  In Bayesian methodology the standard way to compute probability distributions
for unknown parameters is make use of Bayes theorem for the \emph{posterior probability
distribution}
\begin{equation}\label{eq:bayes}
p(\btheta|d, M,I) = \frac{p(d|\btheta, M,I) p(\btheta|M,I)}{p(d|M,I)},
\end{equation}
where $p(d|\btheta, M,I)$ is the \emph{likelihood} of the data $d$ given model $M$ and background
information $I$,  with a set of
parameters $\btheta$, $p(\btheta|M,I)$ is the \emph{prior} on the parameters, and $p(d|M,I)$ is the
\emph{evidence}, or marginal likelihood (in this paper we will use the term evidence throughout
for consistency), of the data given the model. The evidence is the factor that normalises the
posterior probability density. It is given by
\begin{equation}\label{eq:evidence}
p(d|M,I) = \int_{\btheta} p(d|\btheta, M,I) p(\btheta|M,I) \rmd\btheta.
\end{equation}

To compare models, we can calculate the Bayes factor, or {\it odds ratio}, between competing models.
To this end, note that for any model we can calculate its posterior probability as
\begin{equation}
p(M|d,I) = \frac{p(d|M,I)p(M|I)}{p(d|I)}.
\end{equation}
It is hard (maybe impossible) to calculate the normalisation factor $p(d|I)$ as you have to know
{\it all} alternative models and marginalise over them, but we can still compare posterior
probabilities between models provided they use the same data.  We can compute the Bayes factor, or
odds ratio $\mathcal{O}$ (which we will use from here onwards) between two models, as e.g.\
\begin{equation}\label{eq:bayesfactor}
\mathcal{O}_{12} = \frac{p(M_1|d,I)}{p(M_2|d,I)} =
\frac{p(d|M_1,I)}{p(d|M_2,I)}\frac{p(M_1|I)}{p(M_2|I)}.
\end{equation}
Note that the normalising factor $p(d|I)$ has canceled out. If there is no known prior preference
between the two models then the ratio $p(M_1|I)/p(M_2|I)$, the ratio of the prior odds between each
model, can be set equal to unity. In this case, the odds ratio is just the ratio between the
evidences, given by Eqn.~(\ref{eq:evidence}), of the two models. We will adopt this viewpoint
here, so all odds ratios will be calculated as the ratio of evidences.

In the analyses performed in Sections~\ref{sect:waveform_v_source} and \ref{sect:model_selection}
the likelihood function $p(d|\btheta, M,I)$ we use is the Student's $t$ likelihood given in
\citet{Dupuis:2005}. This likelihood assumes that the noise in the data is stationary (over the
defined length of time) and Gaussian, but with an unknown noise standard deviation that has been
analytically marginalised out. However, for the analysis of real data in
Section~\ref{sec:S5results} we have instead estimated the noise level for each data point and therefore
use a Gaussian likelihood function in that section. The reason for this difference in likelihood function
is that for real data it is more efficient to produce our processed data set at a lower sample rate
and with the noise already estimated, which makes the Gaussian likelihood more appropriate.
However, for large numbers of data points the two likelihoods will be very similar.

\subsection{Priors}

To compute evidences and posterior probability distributions we must also explicitly define our
prior probability distributions. For the azimuthal-type angular parameters, and uniform in the
cosine of the polar-type angular parameters, the least informative prior is a uniform prior defined
within their allowed ranges. So, given the ranges in Tables~\ref{tab:priorswaveform} and
\ref{tab:priorssource}, the prior on the angles in the waveform parameterisation, assuming the \tn
model, are

\begin{equation}
p(\Phi_{21}^C, \Phi_{22}^C, \psi, \cos{\iota}|M,I)=
\begin{cases}
\const & \parbox[t]{0.2\textwidth}{if $0 \leq \Phi_{21}^C \leq 2\pi$ \\ and $0 \leq \Phi_{22}^C
\leq 2\pi $ \\ and $0 \leq \psi \leq \pi/2$ \\ and $-1 \leq \cos{\iota} \leq 1$; } \\
0 & \textrm{otherwise,}
\end{cases}
\end{equation}
whilst in the source parameterisation, assuming the \tn model, it is
\begin{equation}
p(\phi_0, \lambda, \psi, \cos{\theta}, \cos{\iota}|M,I)=
\begin{cases}
\const & \parbox[t]{0.2\textwidth}{if $0 \leq \phi_0 \leq 2\pi$\\ and $0 \leq \lambda \leq \pi$ \\ and $0 \leq \cos{\theta} \leq 1$ \\ and $0 \leq \psi \leq \pi/2$ \\ and $-1 \leq \cos{\iota} \leq 1$; } \\
0 & \textrm{otherwise.}
\end{cases}
\end{equation}
Equivalents of these priors for the required parameters are used in the \ta and \biax cases.

We will use priors on the amplitude parameters that are uniform within a range defined
by the
limits in Tables~\ref{tab:priorswaveform} and \ref{tab:priorssource}. These limits on the priors
vary for the different model types, as described in \citet{Jones:2015}.  For the waveform
parameterisation in the \tn case our prior is
\begin{equation}
p(C_{21},C_{22}|M,I)=
\begin{cases}
\const & \parbox[t]{0.3\textwidth}{if $0 \leq C_{21} \leq C_{21}^{\rm max}$
\\ and $0 \leq C_{22} \leq C_{22}^{\rm max}$;} \\
0 & \textrm{otherwise.}
\end{cases}
\end{equation}
but for the \biax case it is
\begin{equation}
p(C_{21},C_{22}|M,I)=
\begin{cases}
\const & \parbox[t]{0.3\textwidth}{if $-C_{21}^{\rm max} \leq C_{21} \leq C_{21}^{\rm max}$\\
and $-C_{22}^{\rm max} \leq C_{22} \leq C_{22}^{\rm max}$\\ and
$C_{22}/C_{21} \geq 0$;} \\
0 & \textrm{otherwise.}
\end{cases}
\end{equation}
For the source parameterisation in the \tn case we choose to use a prior on the amplitudes given by
\begin{equation}
p(I_{31},I_{21}|M,I)=
\begin{cases}
\const & \parbox[t]{0.3\textwidth}{if $0 \leq I_{31} \leq I_{31}^{\rm max}$
\\ and $0 \leq I_{21} \leq I_{21}^{\rm max}$\\ and $I_{31} > I_{21}$;} \\
0 & \textrm{otherwise.}
\end{cases}
\end{equation}
These priors on the amplitude parameters are uniform largely for convenience and simplicity
rather than through a physical motivation. This is consistent with the uniform priors  traditionally
used in searches for \gws from known pulsars, where uniform amplitude priors play a role of allowing a relatively high upper limit to be set
by the likelihood, consistent with the data.
However, when evaluating evidences a choice of uniform
prior does have an influence, as doubling an allowed parameter range
doubles the effective prior volume. However, we limit the effect of this in our analysis by assessing the
distribution of odds ratios between signal and noise models empirically and basing thresholds on
that empirical distribution. Additionally some of the influences of the size of prior volume cancel when comparing signal models.
%
%
It is worthwhile noting that evidence values produced using the minimal parameter ranges given in
Tables~\ref{tab:priorswaveform} and \ref{tab:priorssource} are equivalent to those that would be
produced using the full  parameter space (or, e.g.\ just doubling the $\psi$ ranges). This
is because the likelihood volume within the minimal range is exactly reproduced in each of the
equivalent volumes within the total physical range, along with the prior volume increasing by the
same factor. So, the increase in likelihood volume and prior volume cancel out.

\subsection{Nested sampling}

To calculate odds ratios we need to evaluate the evidence for each model, and Eqn.~(\ref{eq:evidence}) shows this to involve
multi-dimensional integrals. For some parameters,
or likelihoods, the integral may be analytic, or for low numbers of dimensions it may be possible
to evaluate it on a grid, but more generally, efficient numerical integration techniques must be
applied. Here we use the {\it nested sampling} algorithm of \citet{Skilling:2006}, in particular the
implementation of it based on that developed by \citet{Veitch:2010} and available in the {\tt
LALInference} software library \citep{LALInference}. Nested sampling attempts to simplify
Eqn.~(\ref{eq:evidence}) into a one-dimensional integral that can be easily numerically
calculated. It samples a number of {\it live points} from the prior parameter volume, calculates
the likelihood at each point, finds the minimum likelihood $L_{\rm min}$ point to add to the
evidence integral, and then samples a new point with a higher likelihood than $L_{\rm min}$. This
process is repeated until the integral is computed to sufficient accuracy.

The analysis methods and models we have used have been incorporated into a code called
\verb+lalapps_pulsar_parameter_estimation_nested+, which is freely available in the LALSuite
software repository\footnote{\url{
http://www.lsc-group.phys.uwm.edu/daswg/projects/lalsuite.html}}.

\section{Waveform versus source parameters} \label{sect:waveform_v_source}

\begin{figure*}
 \begin{center}
  \includegraphics[width=1.0\textwidth]{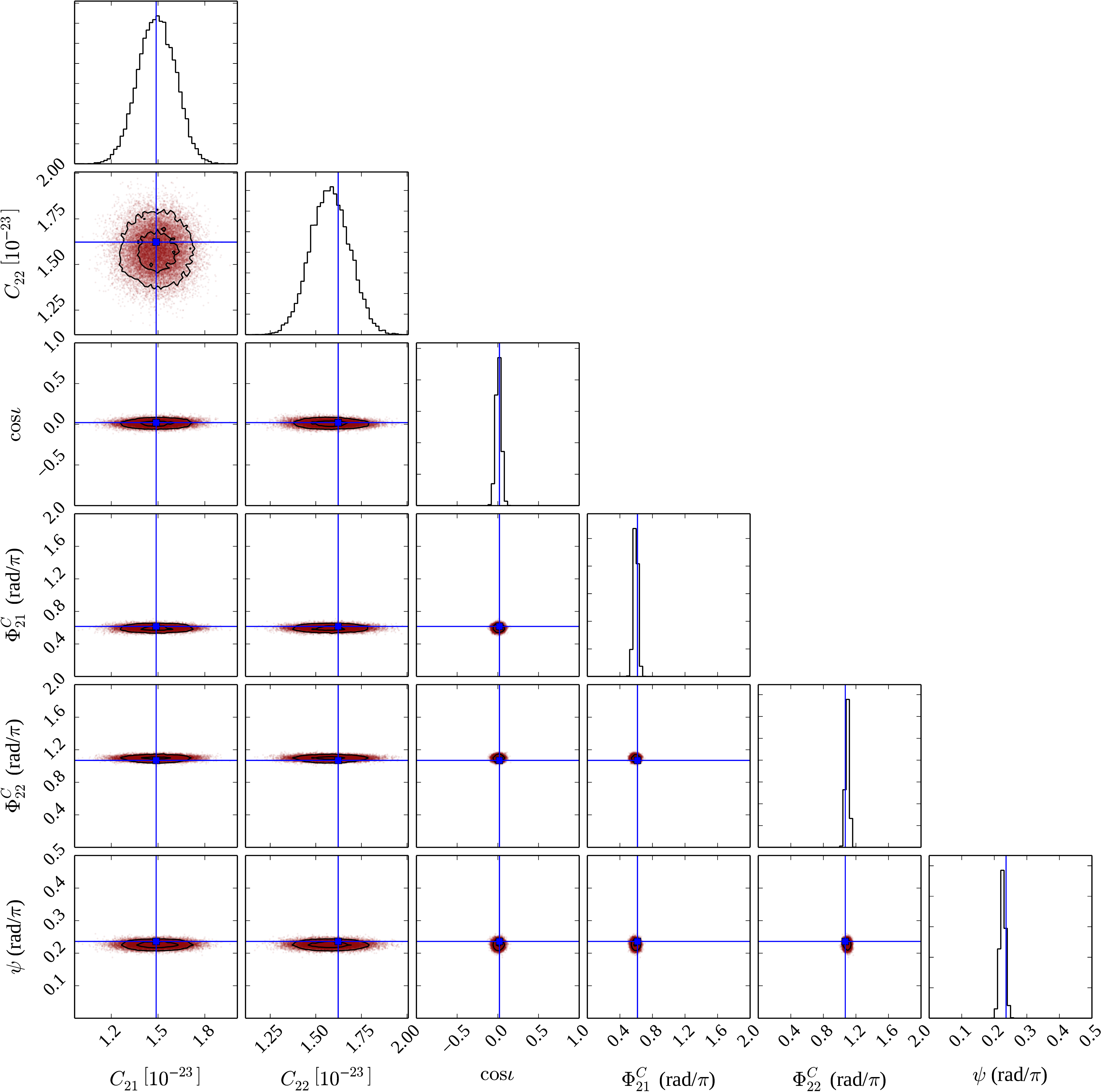}
 \end{center}
 \caption{\label{fig:pdfsignalwaveformlinear} Marginalised posterior probability distribution plots
of waveform parameters for an almost fully linearly polarised ($\cos{\iota} \approx 0$) signal with an
\snr of 20 covering the minimal parameter ranges of Table~\ref{tab:priorswaveform}. The cross-hairs
show the true parameters of the simulated signal. All posterior plots have been produced with a
modified version of the {\tt triangle.py} {\tt python} package
\citep{dan_foreman_mackey_2014_11020}.}
\end{figure*}

\begin{figure*}
 \begin{center}
  \includegraphics[width=1.0\textwidth]{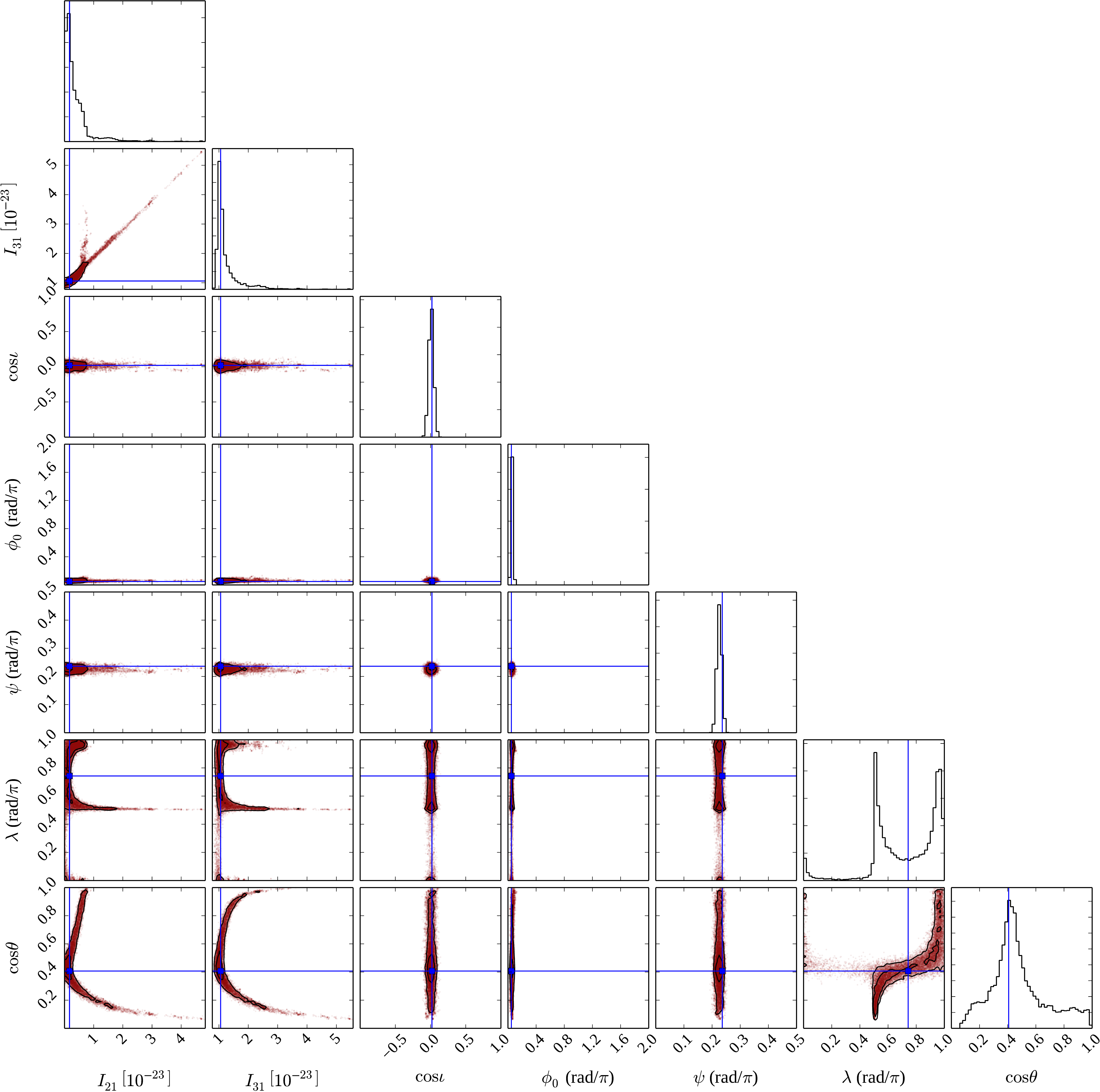}
 \end{center}
 \caption{\label{fig:pdfsignalsourcelinear} Marginalised posterior probability distribution plots of
source parameters for an almost fully linearly polarised ($\cos{\iota} \approx 0$) signal with an \snr
of 20 (the same signal as used in Fig.~\ref{fig:pdfsignalwaveformlinear}) covering the minimal
parameter ranges of Table~\ref{tab:priorssource}. The cross-hairs show the true parameters of the
simulated signal.}
\end{figure*}

\begin{figure*}
 \begin{center}
  \includegraphics[width=1.0\textwidth]{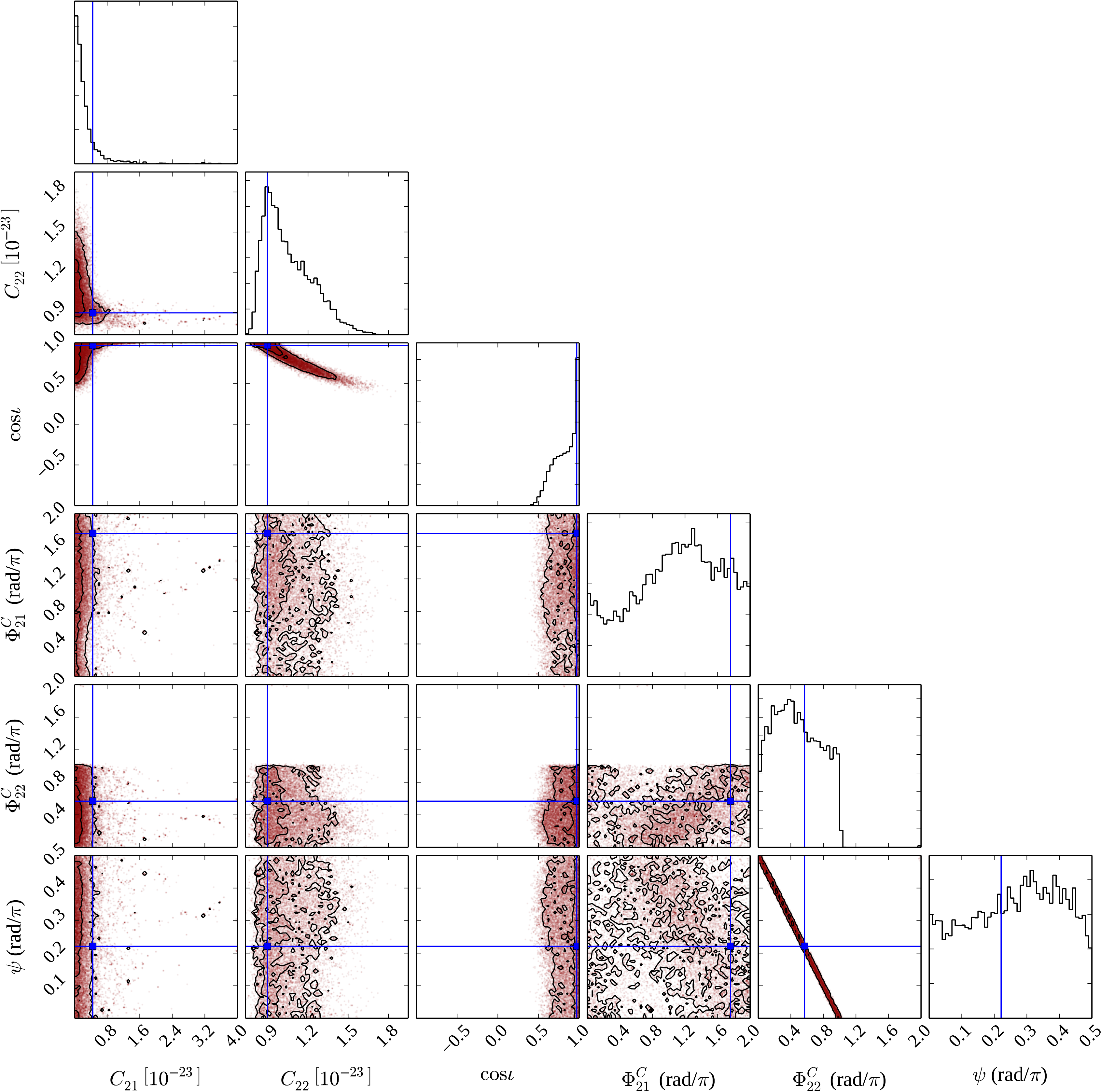}
 \end{center}
 \caption{\label{fig:pdfsignalwaveformcirc} Marginalised posterior probability distribution plots of
waveform parameters for an almost fully circularly polarised ($\cos{\iota} \approx 1$) signal with an
\snr of 20 covering the minimal parameter ranges of Table~\ref{tab:priorswaveform}. The cross-hairs
show the true parameters of the simulated signal.}
\end{figure*}

\begin{figure*}
 \begin{center}
  \includegraphics[width=1.0\textwidth]{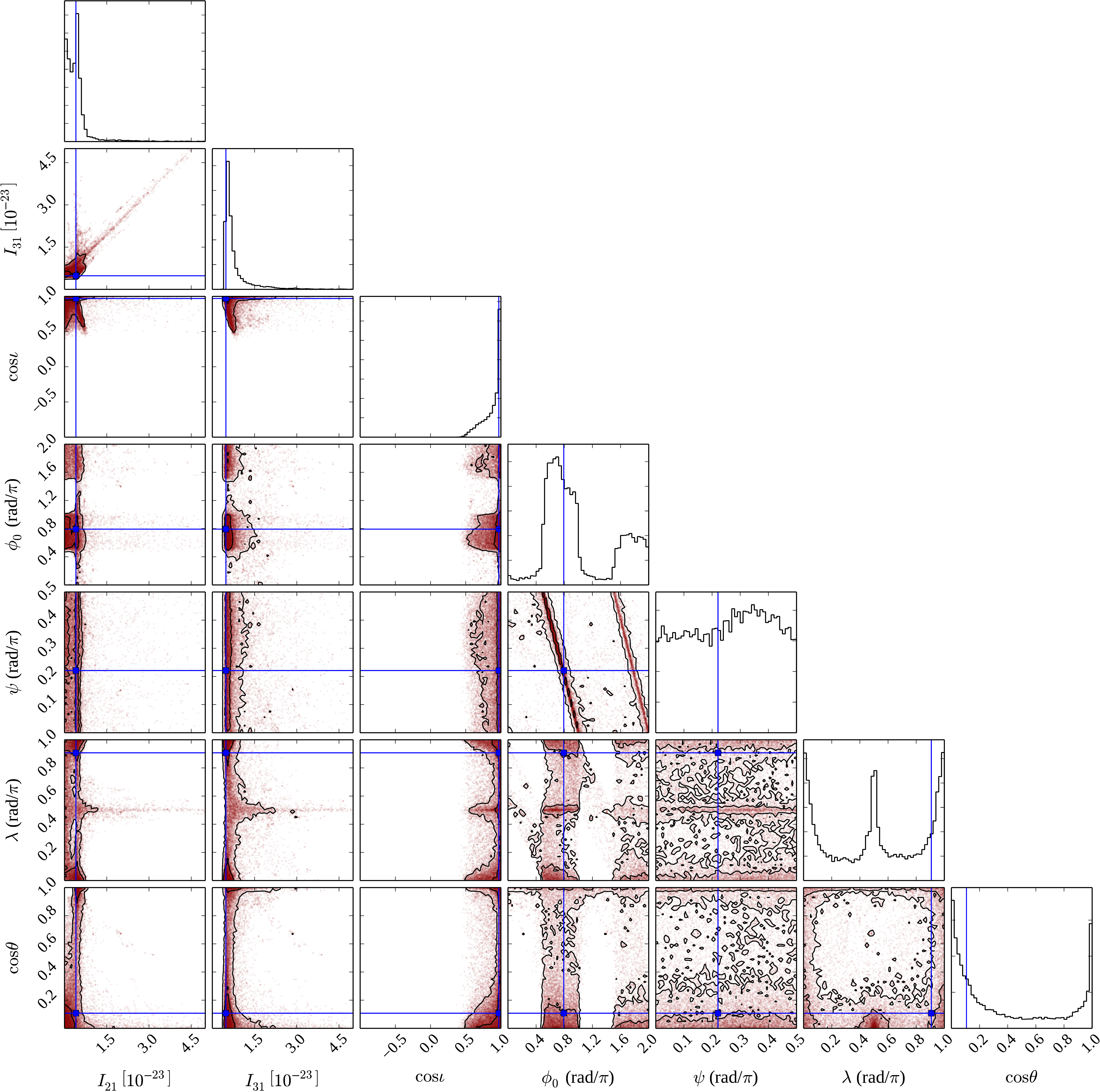}
 \end{center}
 \caption{\label{fig:pdfsignalsourcecirc} Marginalised posterior probability distribution plots of
source parameters for an almost fully circularly polarised ($\cos{\iota} \approx 1$) signal with an
\snr of 20 (the same signal as used in Fig.~\ref{fig:pdfsignalwaveformcirc}) covering the minimal
parameter ranges of Table~\ref{tab:priorssource}. The cross-hairs show the true parameters of the
simulated signal.}
\end{figure*}

It is useful to look at some plots that illustrate the very different nature of the waveform and
source parameters.  To do so, we can make use of the samples produced during nested sampling, by
probabilistically drawing a subset that represent the posterior probability distributions of the
model parameters, using either the waveform or source parameters. The distribution of samples for
an individual parameter (or subset of parameters) represent the posterior probability for that
parameter marginalised over all other parameters. This amounts to integrating
Eqn.~(\ref{eq:bayes}) over the prior ranges given in Tables~\ref{tab:priorswaveform} and
\ref{tab:priorssource} for the required parameter(s). In
Figs.~\ref{fig:pdfsignalwaveformlinear} and \ref{fig:pdfsignalsourcelinear} the one-and-two
dimensional posterior parameter distributions are shown for the \tn model for an almost linearly
polarised signal ($\cos{\iota} \approx 0$) with an \snr of 20, when recovered using the waveform and
source parameters respectively. Equivalent plots for an almost fully circularly polarised signal
($|\cos{\iota}| \approx 1$) are shown in Figs.~\ref{fig:pdfsignalwaveformcirc} and
\ref{fig:pdfsignalsourcecirc}. We present results for these two extremes in inclination angle to
give the reader an idea of the range of different posterior probability distributions than can be
obtained.

From Figs.~\ref{fig:pdfsignalwaveformlinear} and \ref{fig:pdfsignalwaveformcirc} it can be seen
that the waveform parameters show a rather simple uni-modal probability distribution. This is
especially evident for the close-to-linearly-polarised signal, which shows the posteriors to be
largely uncorrelated and Gaussian in appearance; as has been seen in previous \ta analyses, the
extraction of parameters for circular polarisations is slightly more difficult, due to increased
correlations between the parameters \citep{Pitkin:2011}. In contrast, the probability distributions
in the source parameter space, shown in Figs.~\ref{fig:pdfsignalsourcelinear} and
\ref{fig:pdfsignalsourcecirc}, show a large amount of structure \citep[as originally
observed in][]{Gill:2012}.  As described in Appendix~\ref{sect:relating_params}, the five source
parameters $(I_{21}, I_{31}, \theta, \phi_0,
\lambda)$ can be related to the four waveform parameters $(C_{21}, \Phi^{\rm C}_{21}, C_{22},
\Phi^{\rm C}_{22})$.  This leads to the source parameters forming a highly degenerate and curved
tube-like structure, rather than the much simpler form of the waveform parameters. The full complexity
of this tube-like structure is probably being somewhat masked by our choice to only show its
projected marginalisations in two dimensions.

We note that non-negligible probabilities exist out to large values if $I_{21} \approx I_{31}$, due to degeneracies with other
parameters. For example, in Fig.~\ref{fig:pdfsignalsourcecirc}, $I_{21}$ and $I_{31}$ are
truncated to show the bulk of the posterior, but would otherwise show a long tail along the
diagonal of the $I_{21}$ versus $I_{31}$ joint posterior plot. This long tail in $I_{21}$ and
$I_{31}$ is particularly prominent given our choice of a uniform prior in the amplitude parameters.
If we had used a prior uniform in the logarithm of the amplitude parameters then this tail would be
greatly suppressed.

Clearly, it will be much simpler to work with the waveform parameters rather than source
parameters.  There is also the issue of computational speed.  For a stochastic sampling technique
such as nested sampling, the efficiency of the algorithm is greatly increased if new samples can
be drawn from a distribution that closely matches the actual likelihood distribution. If the
true distribution is smoothly varying, uni-modal, and relatively unstructured then it can generally
be well approximated by a multivariate Gaussian. However, for more complex distributions such an
approximation becomes invalid.

Indeed, comparisons in which the analysis code has been run on the same data, but using the
waveform and source parameter spaces, show that to produce a similar number of posterior samples
the former runs $\roughly 1.6$ times faster than the latter in the case of no signal, and $\roughly
1.8$
times faster\footnote{However, this speed difference can
greatly increase when running with larger numbers of live points to get better sampled
posteriors.} for a signal with an \snr of $\roughly 20$. This clearly shows the problems caused when sampling from likelihood functions with
complex structure.

We can therefore see that working with the waveform model rather than the source model is simpler,
both in terms of the dimensionality of the parameter space and the shape of the likelihood
function, and the required computations are faster in the waveform case.  For these reasons, in the
model comparisons that follow in Section~\ref{sect:model_selection}, we will work exclusively in
terms of the waveform parameters.

\subsection{Parameter space mapping}

In the calculations described above, we have used ranges in the parameters that were as small as possible, i.e. we used  the smallest possible sets such that one could be sure that if a triaxial non-aligned signal were present in the data, parameters could be found that matched the signal; see \cite{Jones:2015} for details.  However, in the event of a detection, other parameters can be found that match the signal, in a way described using the transformations given in \cite{Jones:2015}.  Some of these other parameter sets correspond to physically distinct stars.  For instance, if one finds a signal with a polarisation angle $\psi$, there will exist three other solutions with $\psi$ values that differ by successive additions of $\pi/2$, and having different values for some other the other parameters; this degeneracy can only be broken if additional (probably electromagnetic) information is available.  There will also be other parameter sets that correspond to exactly the same physical star, and differ only in a trivial way, relating to how one chooses to label the three Cartesian axes that one lays down on the spinning triaxial body.  It is instructive to fill out the full parameter space, to make clear that these degeneracies exist, and test the transformation rules given in \cite{Jones:2015}.


In the case of the waveform model, we need only enlarge the range covered by the polarisation angle $\psi$, whose  minimal range was $\psi \in [0, \pi/2]$, and so  covered   one quarter of the
full range $\psi \in [0, 2\pi]$ that $\psi$. So, to construct posteriors in the full range we can randomly split the posterior
samples into quarters, and each successive quarter can be mapped into the adjacent parameter volume
by successive application of the transformations given in the Appendix of \citet{Jones:2015}
\begin{align}
 \psi &\rightarrow \psi + \pi/2, \nonumber \\
 \Phi_{21}^C  &\rightarrow \Phi_{21}^C + \pi, \nonumber \\
 \Phi_{22}^C & \rightarrow \Phi_{22}^C + \pi.
\end{align}

For the source model the restricted range only covers a sixteenth of the full range (where
$\cos{\theta} \in [-1, 1]$, $\psi \in [0, 2\pi]$ and $\lambda \in [0, 2\pi]$). This means that the
posterior samples have to be split between the sixteen volumes and more complex transforms used to
map them as given in \citet{Jones:2015}.  If one wishes to map out this full parameter space, considerable care has to be taken when carrying out the
transformations, particularly when selecting the correct roots of inverse trigonometric  functions.
 For this reason, we give in Appendix \ref{sect:algorithm}, a outline of the procedure used here,
written as a simple pseudo-code.

\begin{figure*}
 \begin{center}
  \includegraphics[width=1.0\textwidth]{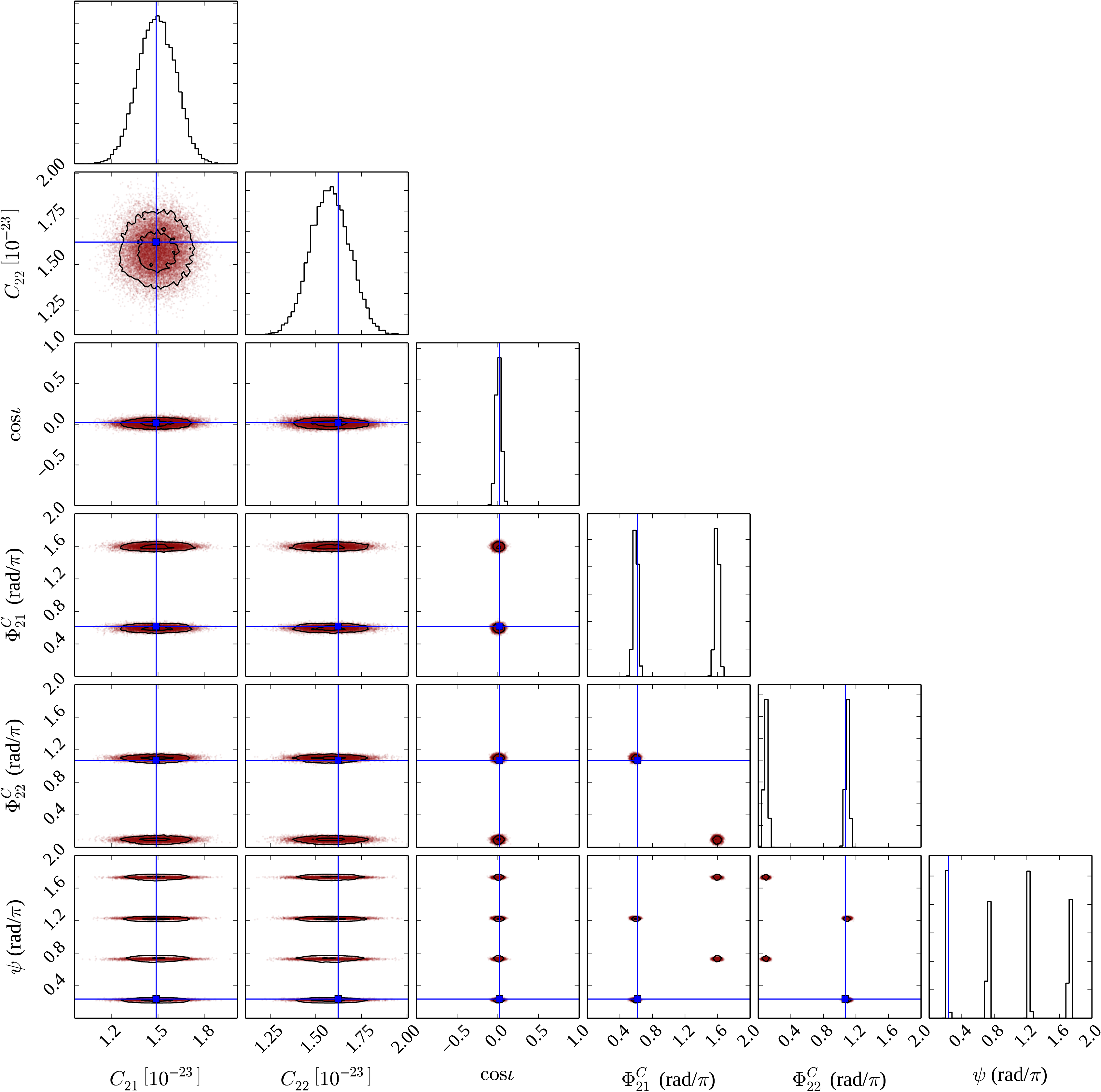}
 \end{center}
 \caption{\label{fig:pdfsignalwaveformlinearfull} Marginalised posterior probability distribution
plots of waveform parameters for the same signal as used in
Fig.~\ref{fig:pdfsignalwaveformlinear}, but covering the full physical parameter ranges. The
cross-hairs show the true parameters of the simulated signal.}
\end{figure*}

\begin{figure*}
 \begin{center}
  \includegraphics[width=1.0\textwidth]{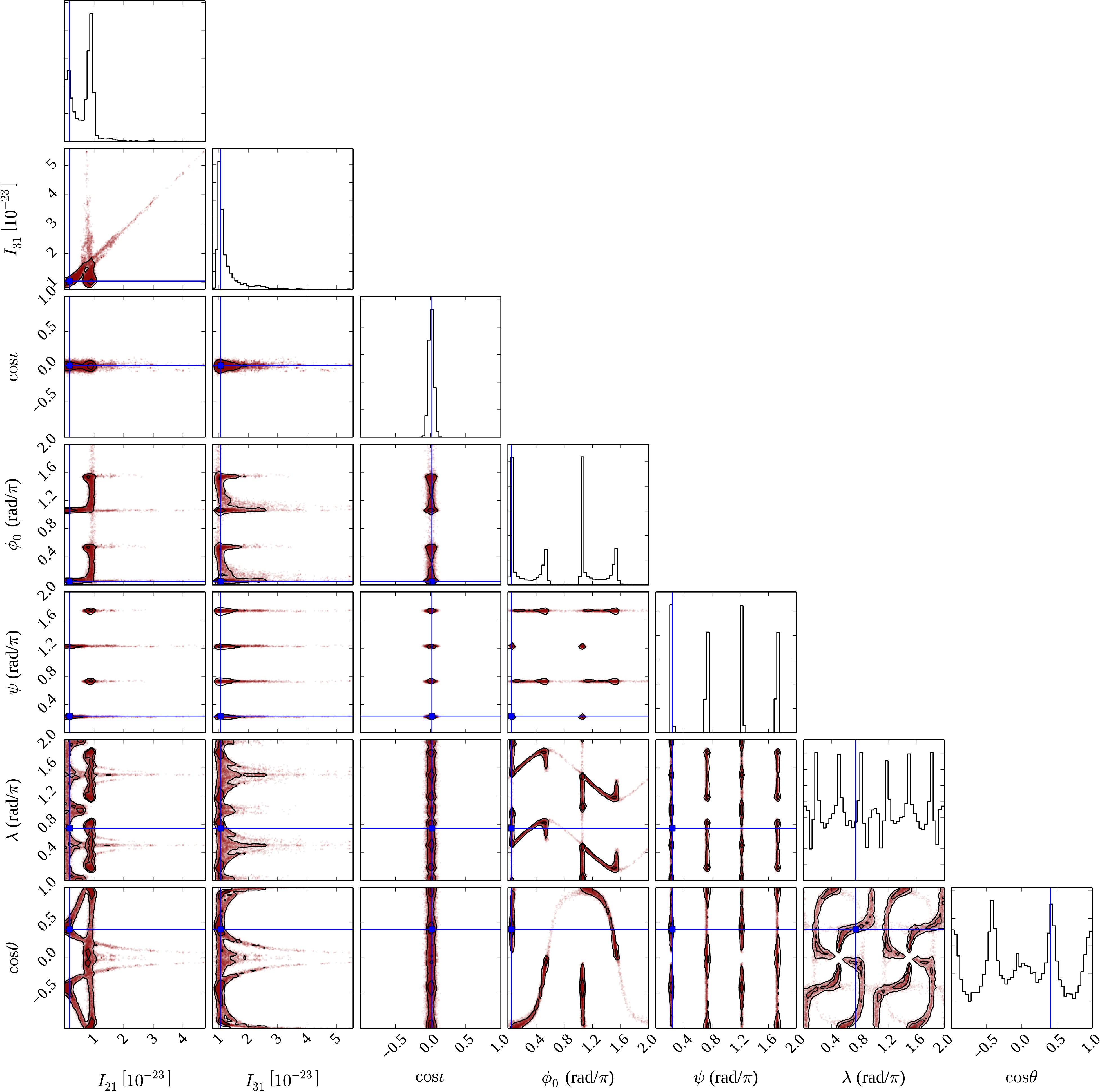}
 \end{center}
 \caption{\label{fig:pdfsignalsourcelinearfull} Marginalised posterior probability distribution
plots of source parameters for the same signal as used in Fig.~\ref{fig:pdfsignalwaveformlinear},
but covering the full physical parameter ranges. The cross-hairs show the true parameters of the
simulated signal.}
\end{figure*}

The full posterior plots for the linearly polarised signal used in
Figs.~\ref{fig:pdfsignalwaveformlinear} and \ref{fig:pdfsignalsourcelinear}, based on this
mapping of posterior samples are shown in Figs.~\ref{fig:pdfsignalwaveformlinearfull} and
\ref{fig:pdfsignalsourcelinearfull}. In the waveform parameterisation shown in
Fig.~\ref{fig:pdfsignalwaveformlinearfull}, whilst the amplitude parameters and inclination can
be unambiguously recovered, it is impossible through the gravitational wave signal alone to be able
to distinguish the combination of initial phases and polarisation angle between the four distinct
modes. In the source parameterisation things become even more complex. Whilst $I_{31}$ and
$\cos{\iota}$ can be reasonably well determined the other parameters suffer from strong
degeneracies. In particular there are always combinations of parameters that allow $I_{21}$ to be
close to zero or large, which means that it may only be possible to ever set upper limits on this
parameter, even in the event of a detection. This implies that for the \ta model precise
determination of the individual physical parameters will not be possible even for high \snr
sources. Only some highly correlated combination of parameters will be precisely determined.

The complex degeneracies of the posteriors for the source parameters show that trying to estimate
parameter uncertainties using the Gaussian approximation of the Fisher matrix would most likely
lead to highly biased results even at high \snrs. However, the waveform parameterisation looks to be
far more amenable to estimation using the Fisher matrix \citep[as is done in][]{Bejger:2014},
provided that the minimal parameter space is used and the multi-modal degeneracies in the full
parameter ranges are subsequently accounted for.

\section{Model selection} \label{sect:model_selection}

In this section we use simulations of signals and noise to evaluate how well we can distinguish
between noise-only data, and data containing a signal of the form of one of our three models (\tn,
\ta, or \biax). In all our simulations we have adopted a noise level for the $f$ and $2f$ data
streams based on the initial LIGO design sensitivity for the 4\,km LIGO Hanford Observatory (H1),
i.e.\ the noise level is not equal between data streams and signal-to-noise ratios will be affected
not just by the signal amplitude, but by their frequency. All simulations assume one day of
heterodyned data sampled at a rate of one sample per minute, as has been standard in previous
searches \citep{Dupuis:2005} and that the evidence evaluation uses a Student's $t$ likelihood
function. For the advanced detectors the sensitivity curve shapes will be similar, but with the
low-frequency edge being pushed to lower frequencies. Note that in all of our analyses, we assume
that the signal phase evolution is tracked exactly, so that the frequency $f$ is a known parameter.

Also in this section we will often refer loosely to the natural logarithm of the odds ratio,
$\ln{\mathcal{O}}$, as the `odds ratio'. This number has the convenience of being far more robust
against issues of numerical precision when dealing with very large or small likelihood values.

\subsection{Noise-only simulations}\label{sec:noiseonly}

The odds ratio itself tells you how much one model is favoured over another, but computed odds ratios
will be numerically different for different noise realisations. \citet{Jeffreys:1931} gives a rule-of-thumb table
for assessing the significance of odds ratios, however this can also be approached empirically by using
simulations to determine their distribution over realisations. In particular, by assessing the odds
ratios empirically we can alleviate the effect of our large prior amplitude volume in
shifting the distribution of odds ratios towards low values even when signals could
potentially be seen. We will therefore follow this second path. The dependence of the odds ratios
we calculate on the prior volume of the amplitude parameters probably shows that our choice of
amplitude priors (uniform up to some maximum, and zero above the maximum) could be improved upon.  One alternative option is a prior that
is uniform up to a lower maximum value and then uniform in the logarithm of amplitude for larger
values.   However, the simple choice we have made has the advantage that the same prior can be used for all
sources, no matter where they sit within the detector's sensitivity curve, with the knowledge that
the prior will easily bound the bulk of the likelihood in all cases.

\begin{figure}
 \begin{center}
  \includegraphics[width=0.49\textwidth]{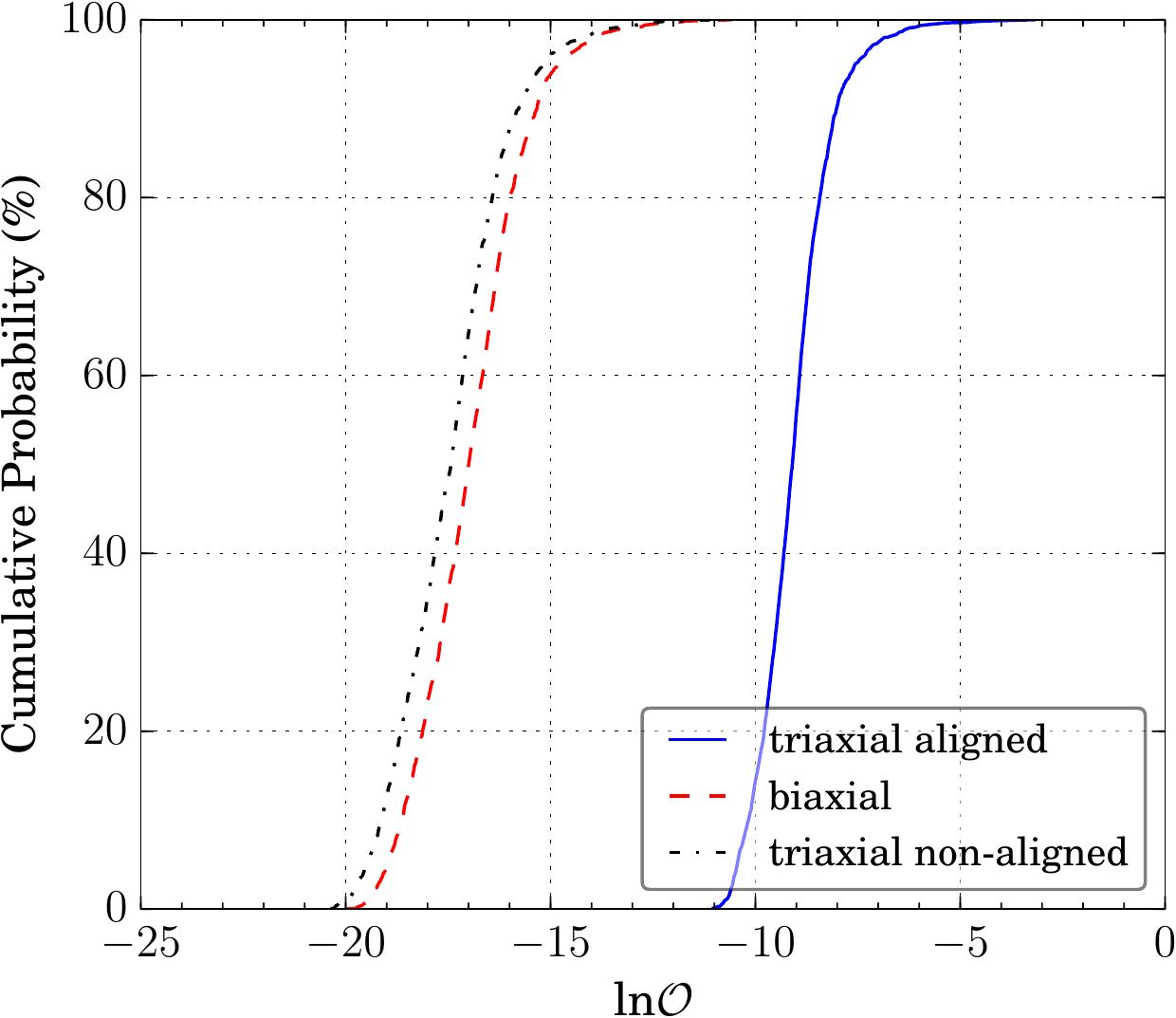}
 \end{center}
 \caption{\label{fig:noisedist} The cumulative probability distribution of odds ratios comparing
the three signal models to Gaussian noise for data containing only noise.}
\end{figure}

For our noise-only analysis we ran 2\,000 simulations, with $f$ drawn from a uniform distribution
between 50 and 700\,Hz, and the $f$ and $2f$ data streams generated by drawing the real and
imaginary components from Gaussian distributions with zero mean and a variance set by the
expected H1 noise at the given frequency\footnote{The variance is calculated as $\sigma(t)^2 =
S_n(f)/4\Delta{}t$, with $S_n(f)$ being the one-sided power spectral density at $f$ and $\Delta{}t$
being the time series time step, which is 60\,s for this analysis.}. For each simulation a
random sky location was chosen, generated from a uniform distribution on the sky, as this
determines the antenna pattern functions $F_+$ and $F_\times$ of Eqns~(\ref{eq:hetf}) and
(\ref{eq:het2f}). Note that, for the reasons outlined previously, when calculating odds ratios, we
used signals written in terms of the waveform parameters rather than source parameters. For each
simulation the odds ratios of evidence for each model versus that for Gaussian
noise\footnote{For Gaussian noise the evidence is calculated by setting the signal amplitude to
zero in the likelihood function.} have been calculated using nested sampling, and the cumulative
probability distributions of all these values are shown in Fig.~\ref{fig:noisedist}. Note that for
the \ta model only the $2f$ data stream was used.

Fig.~\ref{fig:noisedist} has three notable features.  First, we see that the odds ratio
($\rme^{\ln{\mathcal{O}}}$) for all three signal models is, for a large fraction of the simulations,
small, confirming that noise-only data is normally found to be more consistent with noise than with
a signal, as expected. Second, for a given cumulative probability, the odds ratios for the \biax
and \tn models are much smaller than for the \ta model. There are two related explanations for
this: it is intrinsically less likely for noise to conspire to imitate a signal with components at
both $f$ and $2f$, as this would require the noise to produce signal-like disturbances at the two
widely-separated frequencies, with correlated properties (e.g.\ with consistent values for the
parameters $\iota$ and $\psi$); and, there will be an Occam factor at play, such that in the
absence of any evidence for a particular signal the simpler \ta model, which has fewer parameters
and a correspondingly smaller prior volume than the \biax and \tn models, will be favoured. In
particular the large extra prior volume added by the extra amplitude parameter in the \tn and \biax
models play the predominant role in the shift between the odds ratio distributions. Third, the
curves show that the \biax model is slightly favoured over the \tn model. This again can be accounted for
through the Occam factor, due to the additional parameter, and therefore prior volume, required for
the \tn model over the \biax model.

From these distributions we can set an odds ratio threshold at which we favour one model over noise
at a given false alarm probability. If we choose a false alarm probability of 1\% we find threshold
odds ratios for each model versus Gaussian noise alone are $-6.3$, $-13.0$, and $-13.5$ for the
\ta, \biax and \tn cases respectively. We will use these thresholds for calculating detection
efficiencies below. However, it is worth noting that analyses of data of different lengths of time,
and/or combining additional detectors, would produce different values for the odds ratios. This
means that a threshold needs to be calculated for a specific analysis and that the values above are
only relevant if using one day of LIGO H1 data sampled at a rate of once per minute. A different threshold
would be required for the analysis of real LIGO data presented in Section~\ref{sec:S5results}, and
indeed in Section~\ref{sec:significance} we demonstrate a different, but related, assessment of
detection significance. It is also worth noting that these simulations have used Gaussian noise,
whereas the distribution of odds ratios for real data would most likely be different.

\subsection{Signal simulations}\label{sec:signalinj}

To assess detection efficiencies for signals described by the three different models we have
generated simulations including these signals.  In all three cases we drew the signal sky positions
from a uniform distribution on the sky, and chose amplitude parameters to give a uniform
\snr distribution between $0$ and $20$. We chose the distribution of angular parameters to be
uniform within the prior ranges given in Table~\ref{tab:priorssource}. In the \tn case we drew
values from the source parameters, with $I_{21}$ drawn from a uniform distribution between zero and
an upper range, whilst for each $I_{21}$ value $I_{31}$ was drawn from a uniform distribution
between $I_{21}$ and the same upper range, thus ensuring that $I_{31} > I_{21}$. To obtain a
uniform distribution in the overall \snr for the combined $f$ and $2f$ signal each of these pairs
of parameter was re-scaled such that the final distribution was uniform in \snr between $0$ and $20$.

In this analysis we define the \snr for a discretely-sampled complex signal, $x_i$, as
\begin{equation}
 \rho = \frac{1}{\sigma}\left\{\sum_{i=1}^N \left[\Re{(x_i)}^2 + \Im{(x_i)}^2\right]\right\}^{1/2}
\end{equation}
where $\sigma$ is the noise standard deviation (assumed constant) in both the real and imaginary components of the data,
The coherent \snr for a combined $f$ and $2f$ signal is
then just given by $\rho_{\rm coh} = (\rho_f^2 + \rho_{2f}^2)^{1/2}$.

We generated a set of 2\,000 simulations including signals from the full \tn model, 2\,000
simulations including signals from the \ta model, and 2\,000 simulations including signals from the
\biax model, with coherent \snrs between 0 and 20. To assess model comparison at much higher
\snrs we generated  further sets of 500 injections of \tn signals with \snrs of 50, 100 and 500.

For each set of injections we calculated the odds ratios of signal vs.\ Gaussian noise
for each of the three different models. We then used these to compute detection
efficiencies, and also to compute odds ratios between the three different signal models. We now
present the results for each type of injected signal.

\subsubsection{Detecting \tn signals}

Using the odds ratio threshold for the 1\% false alarm probability found in
Section~\ref{sec:noiseonly}, we can work out the efficiencies of each model for detecting an
injected \tn signal.  We can do this as a function of the total \snr, but also as a function
of the \snr in both the $f$ and $2f$ streams individually. We can also use these odds ratios to compare the
signal models against one another.

\begin{figure}
 \centering

 \subcaptionbox{\label{subfig:efffull}}{
 \includegraphics[width=0.425\textwidth]{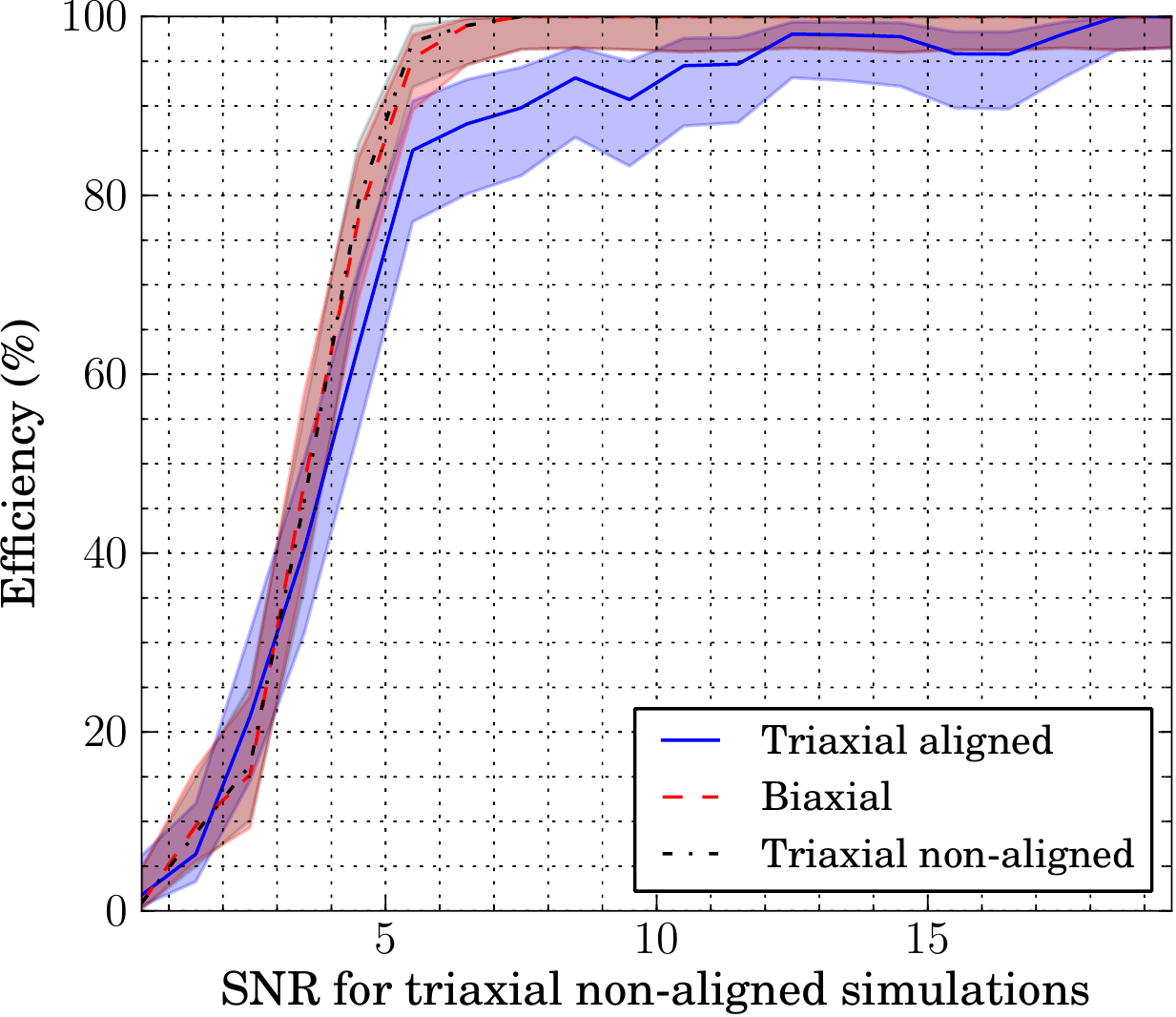}}
 \smallskip

 \subcaptionbox{\label{subfig:efftri}}{
 \includegraphics[width=0.425\textwidth]{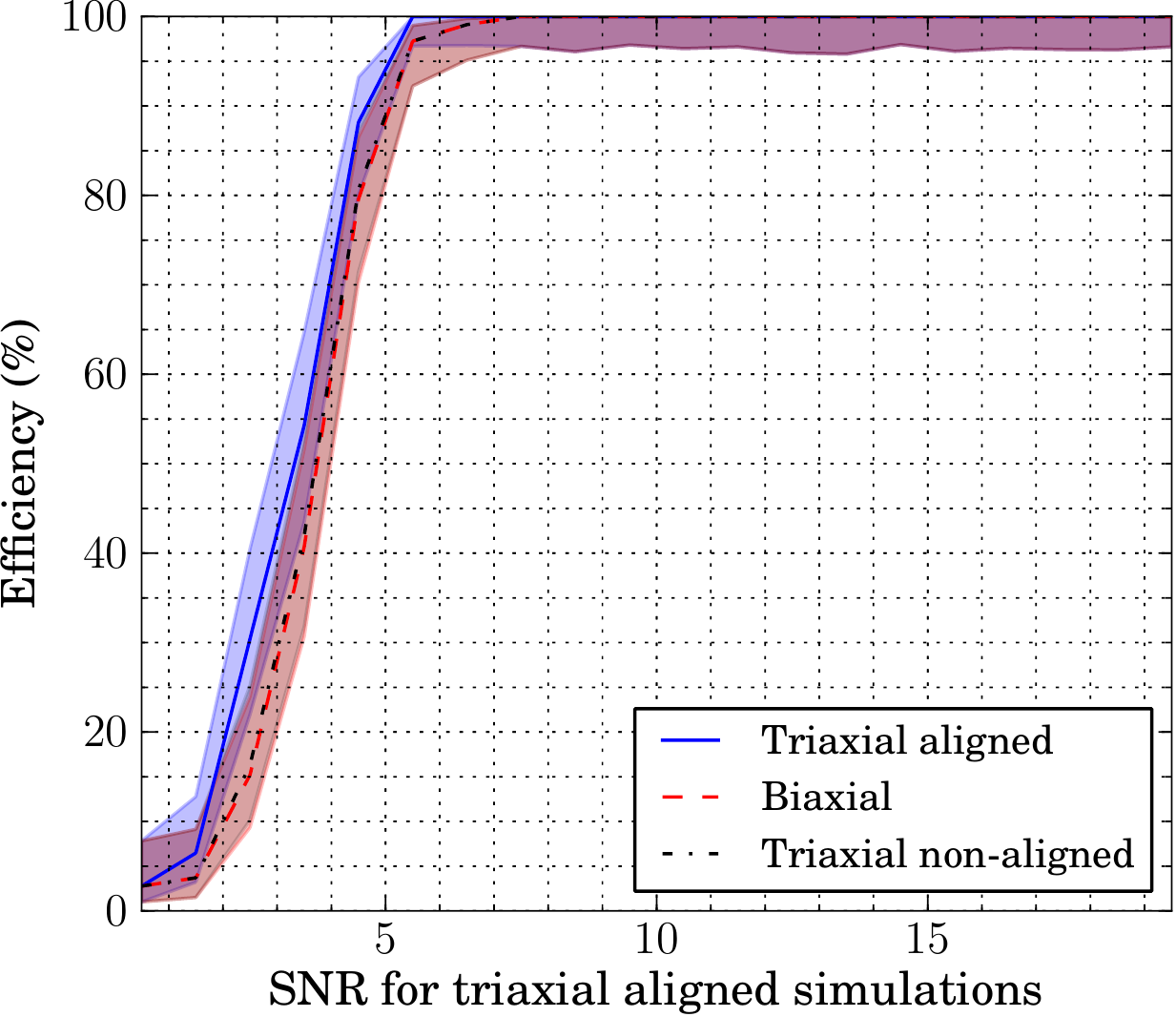}}

 \smallskip
 \subcaptionbox{\label{subfig:effbi}}{
 \includegraphics[width=0.425\textwidth]{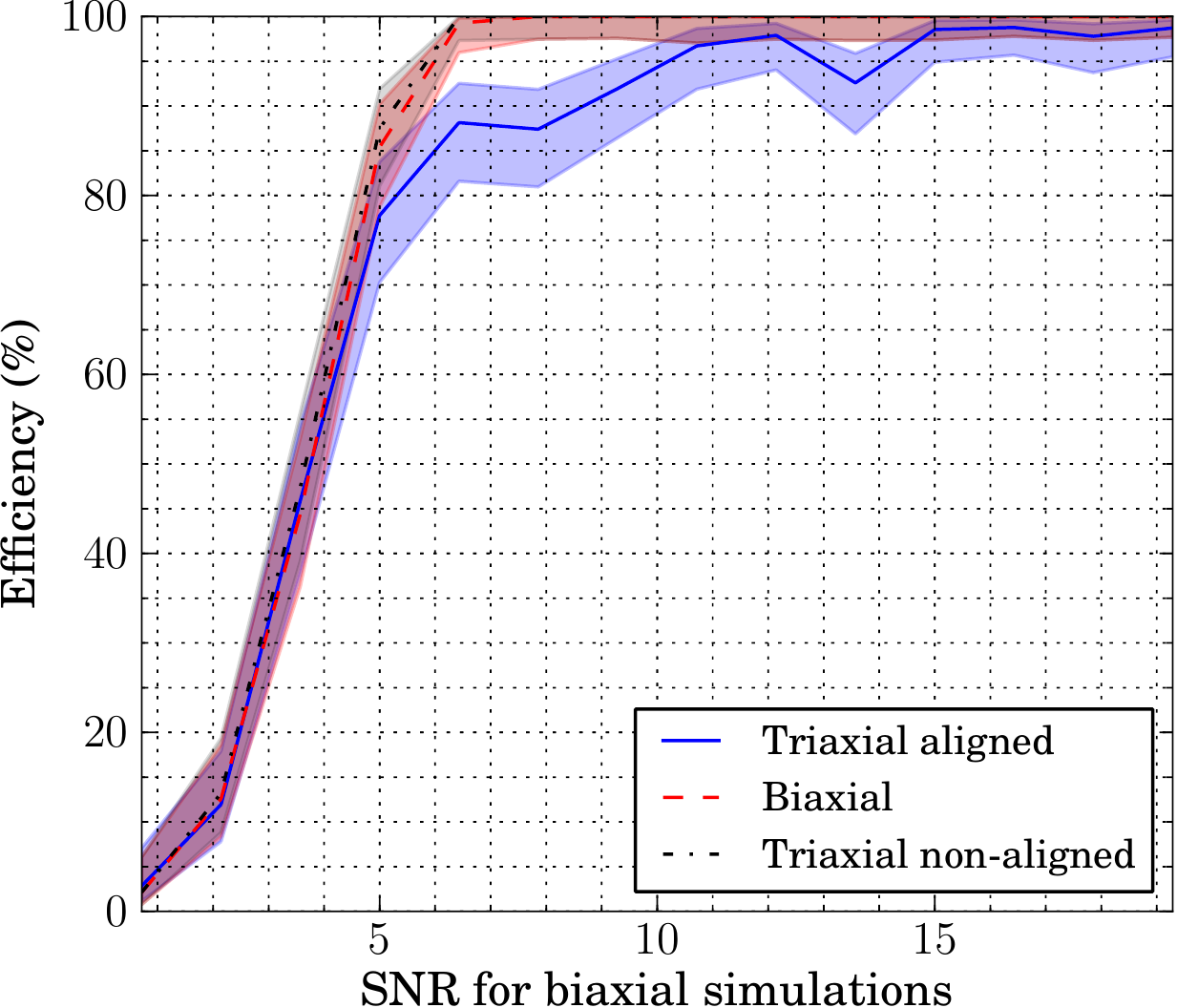}}

\caption{\label{fig:effvstotsnr} Each figure shows the efficiencies for signal detection (with a
1\% false alarm probability) based on the odds ratios recovered assuming the three different signal
models.  In (\subref{subfig:efffull}) the data contained \tn injections, in (\subref{subfig:efftri})
\ta injections, and in (\subref{subfig:effbi}) \biax injections.}
\end{figure}

\begin{figure}
 \centering
 \subcaptionbox{
 \label{subfig:triaxialn}}{
 \includegraphics[width=0.41\textwidth]{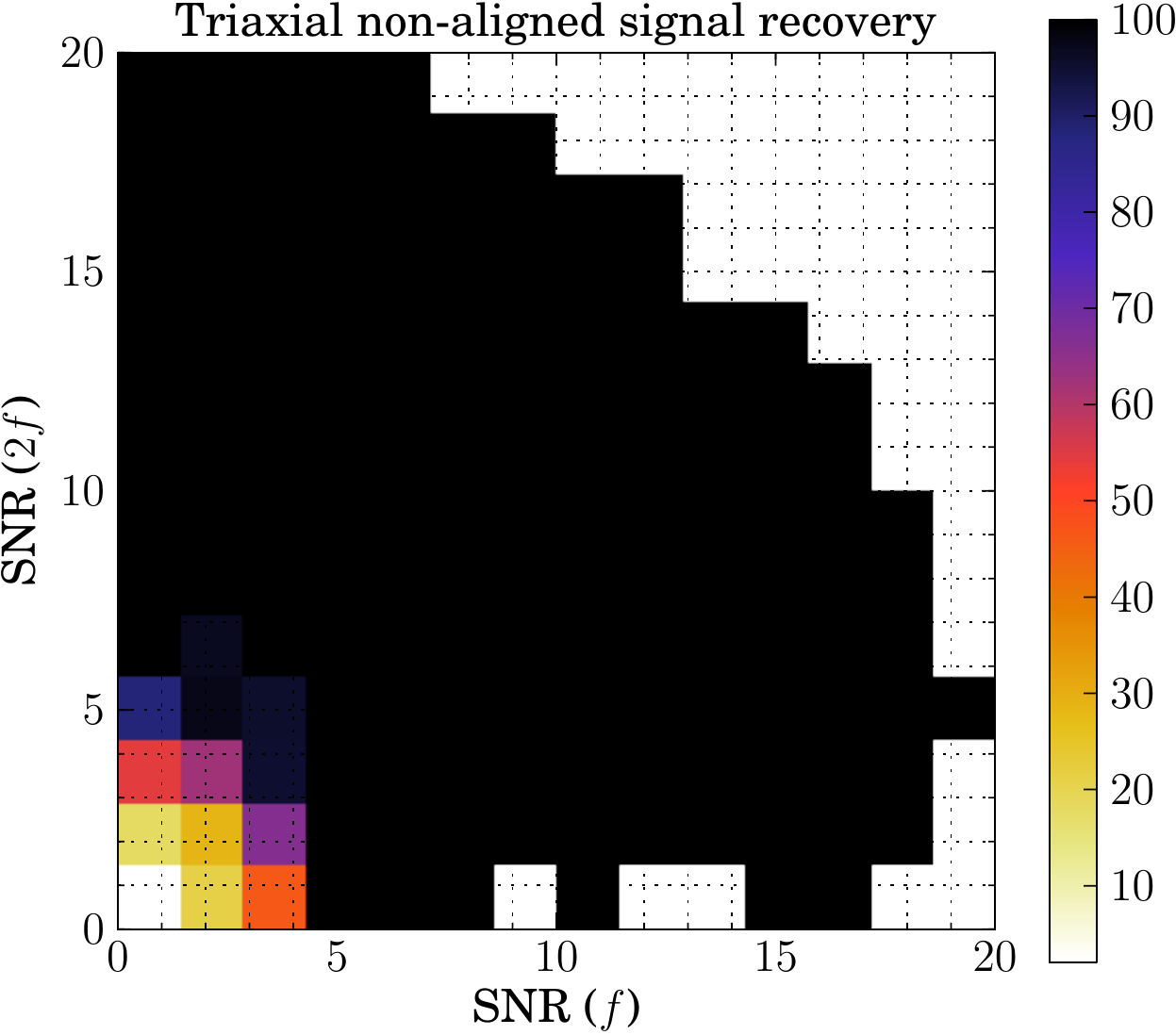}}
 \smallskip

 \subcaptionbox{\label{subfig:triaxial}}{
 \includegraphics[width=0.41\textwidth]{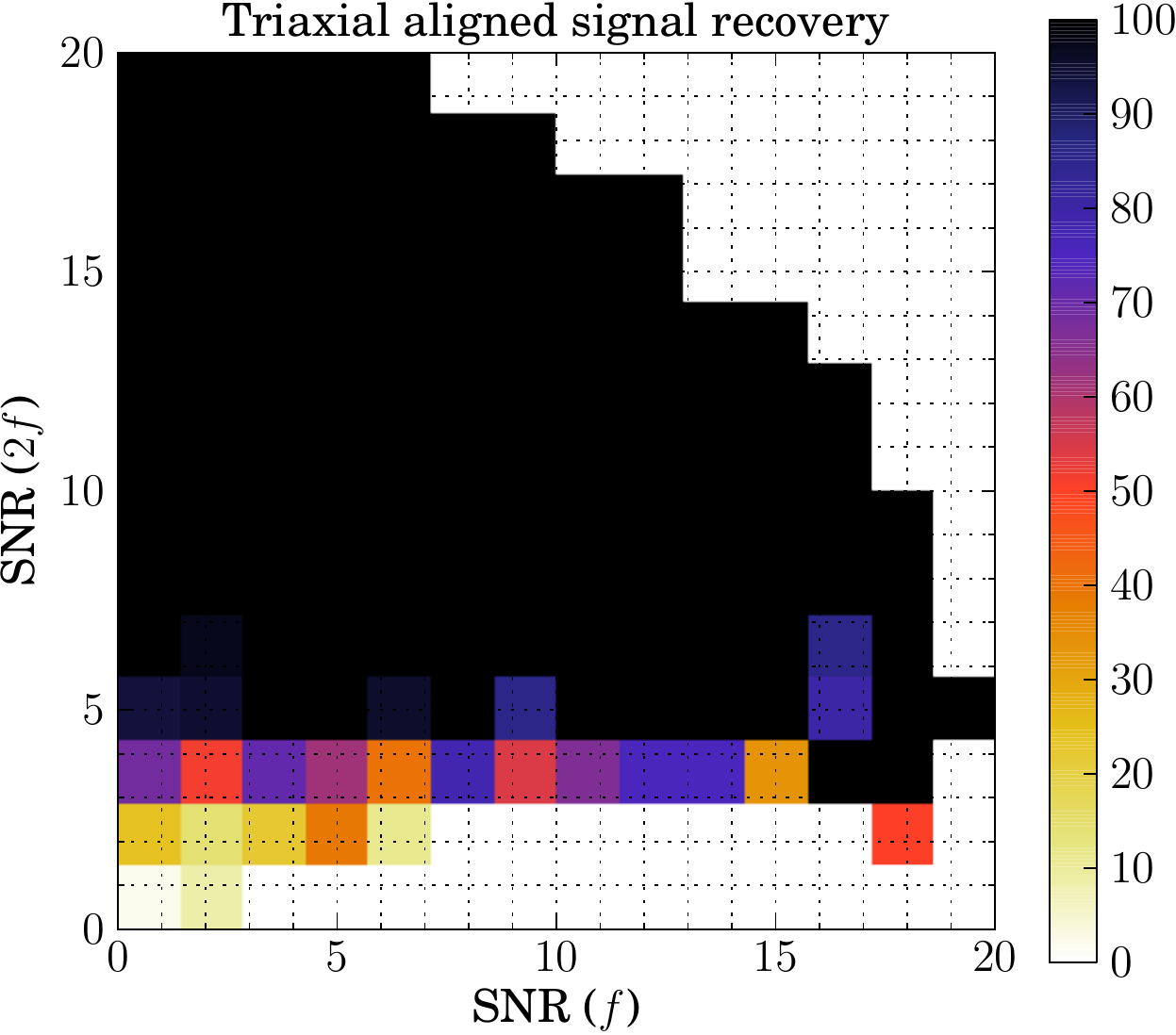}} 
 \smallskip
\subcaptionbox{
\label{subfig:biaxial}}{
 \includegraphics[width=0.41\textwidth]{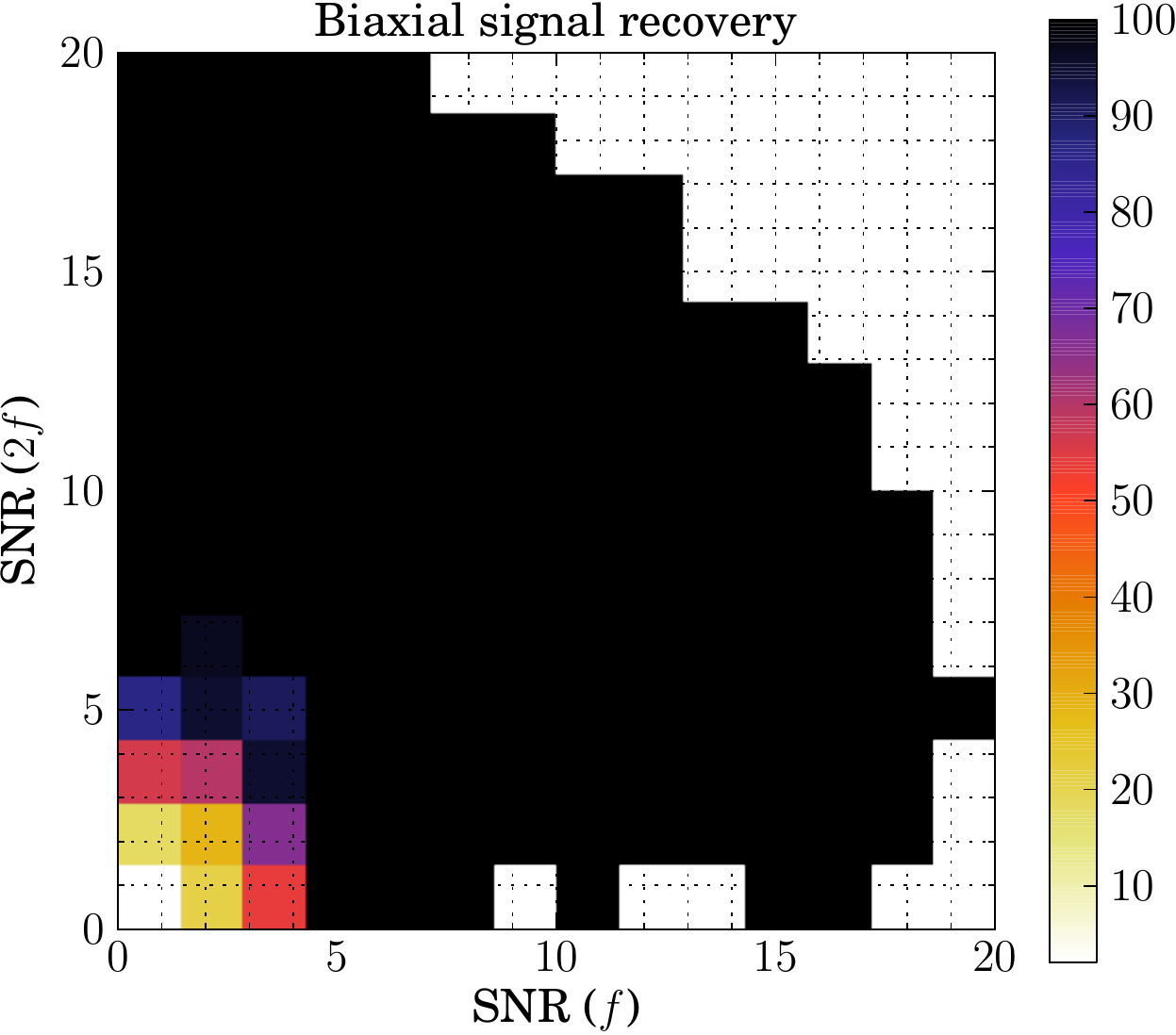}}

\caption{\label{fig:effvssnrs} The efficiencies
for signal detection (with a 1\% false alarm probability) based on odds ratios for data containing
\tn injections, plotted as a function of the signal \snr in the individual $f$ and $2f$ data
streams. The analysis in (\subref{subfig:triaxialn}) assumed the (correct) \tn model, while the
analyses in (\subref{subfig:triaxial}) and (\subref{subfig:biaxial}) assumed (incorrectly) the \ta and
\biax models, respectively.}
\end{figure}

Fig.~\ref{subfig:efffull} shows the efficiency for detecting \tn signals as a function of the
combined \snrs in the $f$ and $2f$ data streams \citep[the shaded regions give 95\% credible intervals
calculated using the method of][]{Cameron:2011}.The three curves correspond to using
the (correct) \tn model, and using the (incorrect) \ta and \biax models.  The equivalent
efficiencies as a function of the \snr in each individual stream are shown in
Fig.~\ref{fig:effvssnrs}, with Fig.~\ref{subfig:triaxialn} assuming the (correct) \tn model,
and Figs.~\ref{subfig:triaxial} and \ref{subfig:biaxial} assuming the (incorrect) \ta and \biax
models, respectively.

From Fig.~\ref{subfig:efffull} it can be seen that a 95\% detection
efficiency is achieved for the \biax and \tn models for \snrs $\gtrsim 5.5$. The
detection efficiency using the \ta model is systematically lower. The \biax and \tn models
recover signals with nearly equal efficiency showing that even though the injected \tn signal
contains an additional phase parameter, recovery with the \biax model generally finds a nearly
equivalent parameterisation to match it (see below). It is obvious that the \ta model will not detect signals
for which there is very little power in the $2f$ stream, which is the reason for its detection
efficiency curve in Fig.~\ref{subfig:efffull} always being slightly lower than that of the
triaxial non-aligned and biaxial models.  This is seen as an obvious effect in
Fig.~\ref{subfig:triaxial}.  However, signals with \snr $\gtrsim 6$ in the $2f$ stream are still
well recovered. (In contrast, when assuming a \tn or \biax signal, detection is possible providing the \emph{combined} SNR is sufficiently large, as made clear in
Figs.~\ref{subfig:triaxialn}  and \ref{subfig:biaxial}, respectively).  This again shows that, even for signals with the full \tn model parameterisation, the
$2f$ component can still be well-matched with the \ta model. This good match is straightforwardly
apparent when thinking in terms of the waveform parameterisation as the signal in the $2f$
component has exactly the same form for both models, whereas
this equivalence is not so evident when using the source parameterisation.

\begin{figure*}
\centering

\subcaptionbox{\label{subfig:modcompfull}}{
 \includegraphics[width=0.6\textwidth]{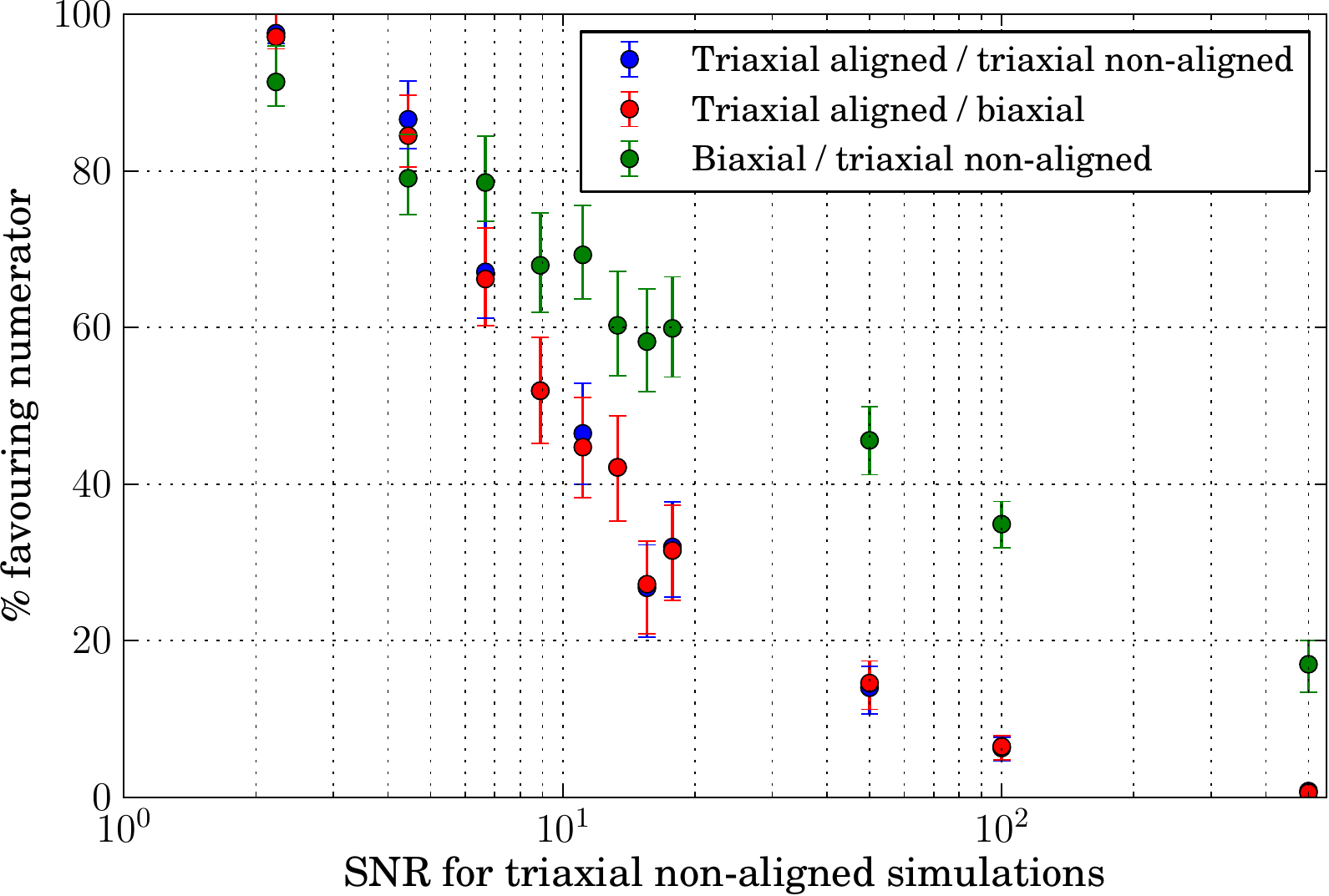}}
\smallskip

\subcaptionbox{\label{subfig:modcomptri}}{
 \includegraphics[width=0.425\textwidth]{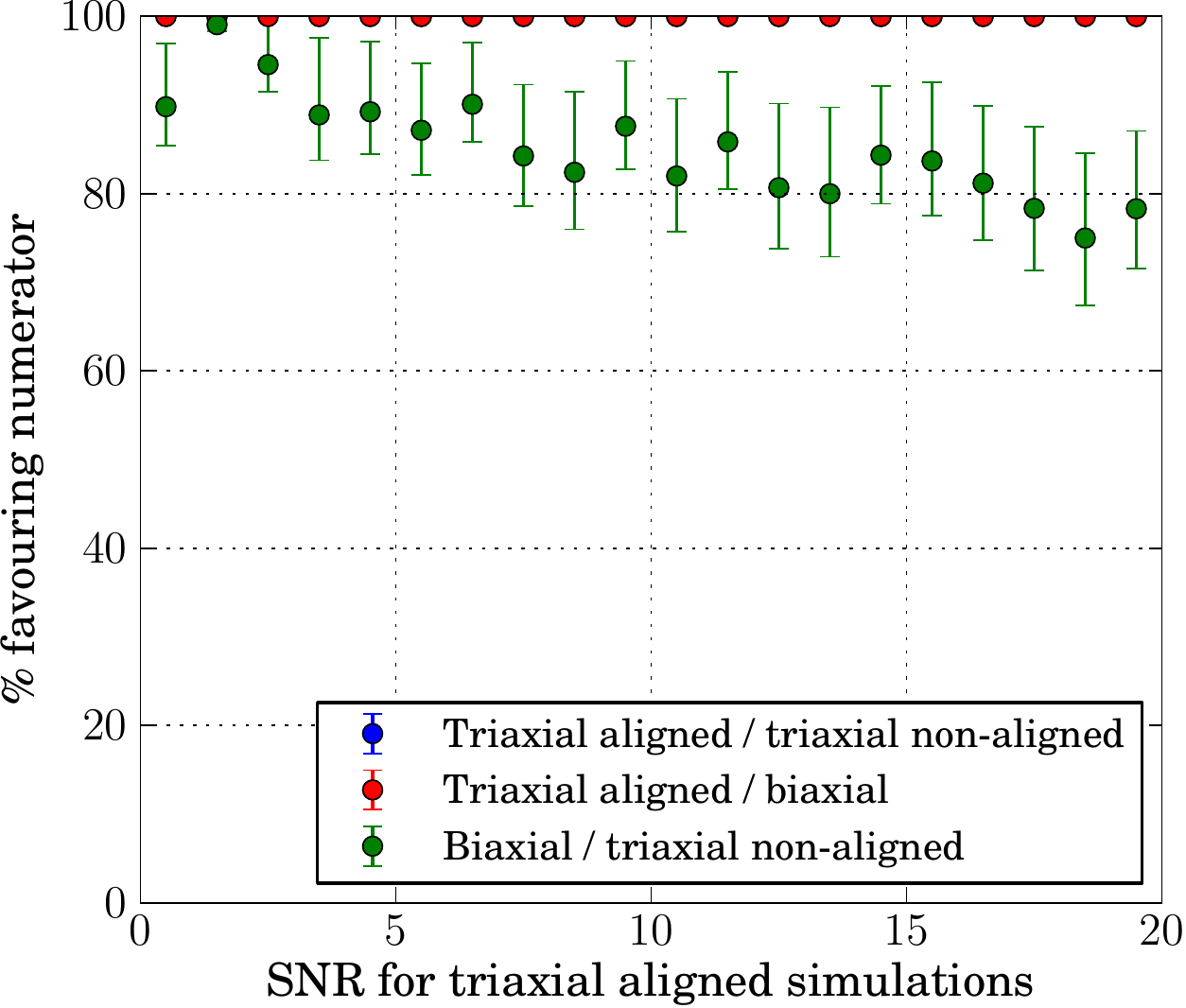}}
\hspace{1em}
\subcaptionbox{\label{subfig:modcompbi}}{
 \includegraphics[width=0.425\textwidth]{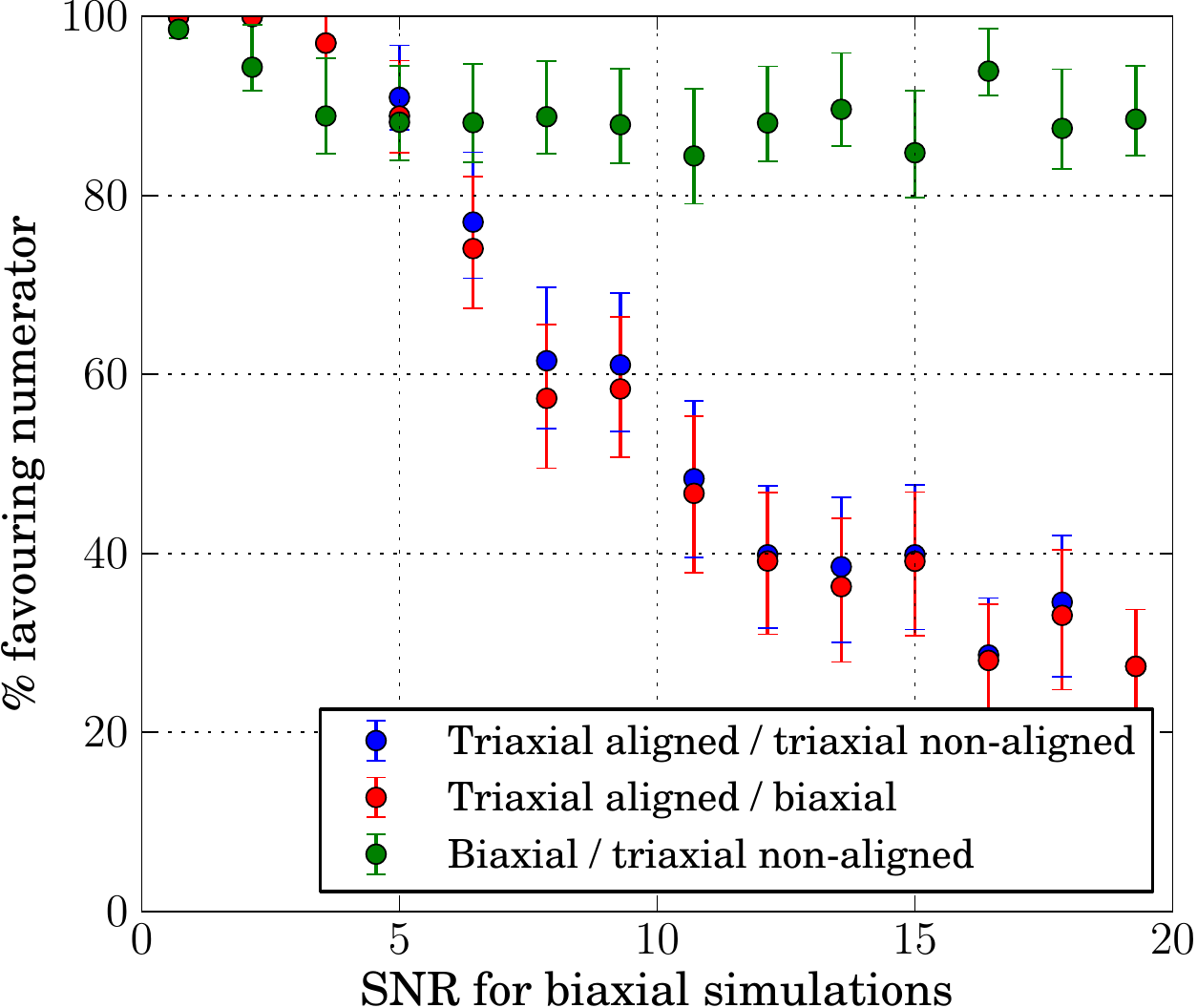}}
\caption{\label{fig:modcomp} The percentage of simulations favouring a particular model (see legends)
when the simulation contains (\subref{subfig:modcompfull}) the full \tn model,
(\subref{subfig:modcomptri}) the \ta model, and (\subref{subfig:modcompbi}) the \biax model, as a
function of the signal coherent \snr.   Note that in (\subref{subfig:modcomptri}) the (blue) \ta/\tn results are barely visible beneath the (red) \ta/\biax results.}
\end{figure*}

Using the odds ratio values we have calculated, we are also able to compare the Bayesian evidences
between signal models rather than just comparing a model to noise. For the \biax and \tn cases we
can take the ratio of odds ratios that we have already calculated as the noise evidence terms will
cancel out. However, for the \ta case we need to include the evidence for there being no signal
present in the $f$ data stream as part of the signal model. Therefore the evidence for the \ta
model becomes the product of the evidence for a signal at $2f$ and only noise at $f$. We take
model 1 to be favoured over model 2 if $\mathcal{O}_{12} \equiv \mathcal{O}_1/\mathcal{O}_2 > 1$. For
each model pairing (\ta versus \tn, \ta versus \biax and \biax versus \tn) we have computed the
percentage of our simulated signals for which the numerator model is favoured as a function of \snr,
e.g.\ 50\% means that on average either model is equally likely. Results for this can be seen in
Fig.~\ref{subfig:modcompfull}.

\begin{figure*}
 \begin{center}
  \includegraphics[width=1.0\textwidth]{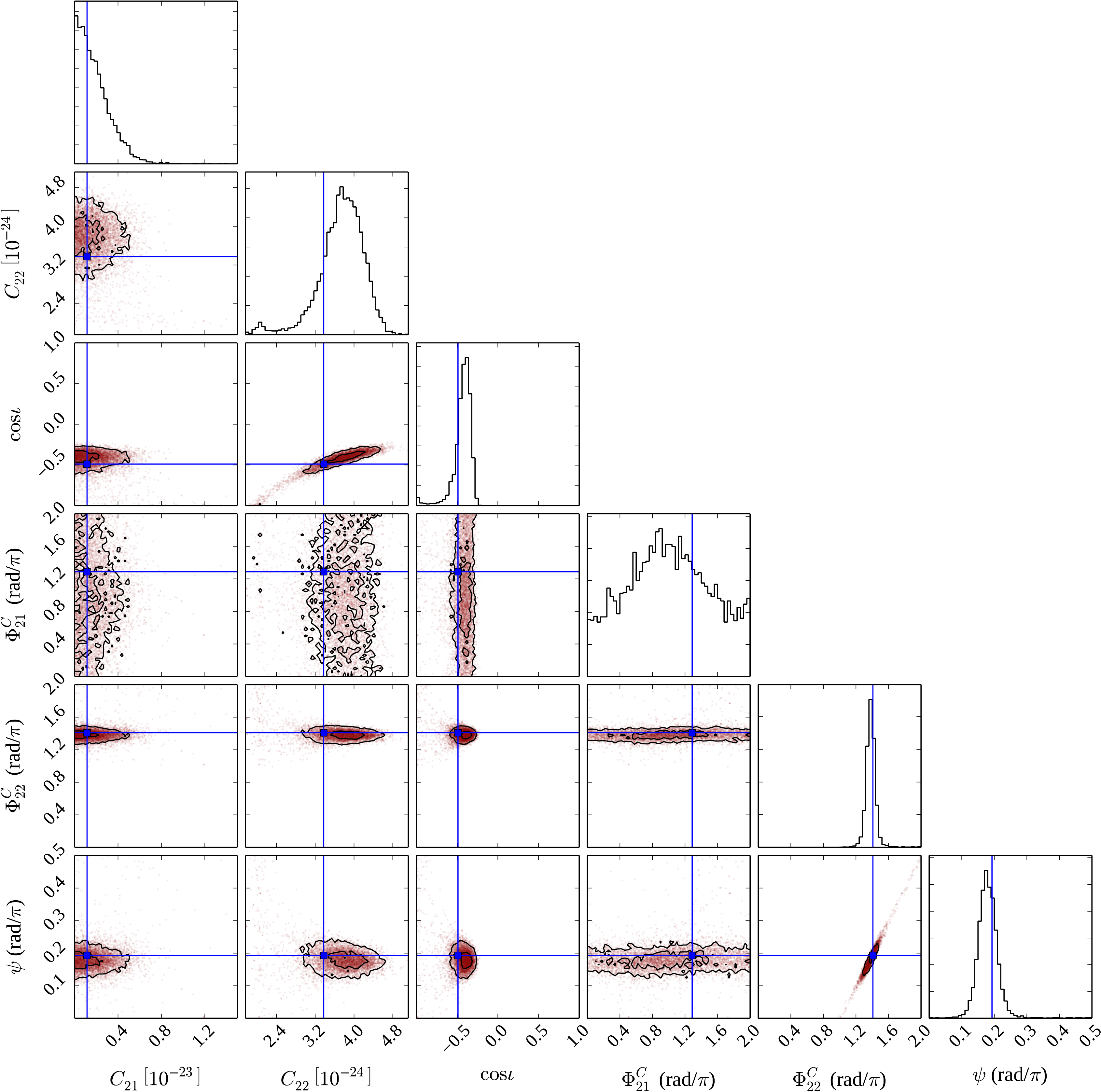}
 \end{center}
 \caption{\label{fig:cosiota} Marginalised posterior probability distribution plots
of waveform parameters for an \snr $\roughly 20$ signal with $\cos{\iota} = -0.49 $ covering
the minimal parameter ranges of Table~\ref{tab:priorswaveform}.}
\end{figure*}

\begin{figure}
 \begin{center}
  \includegraphics[width=0.47\textwidth]{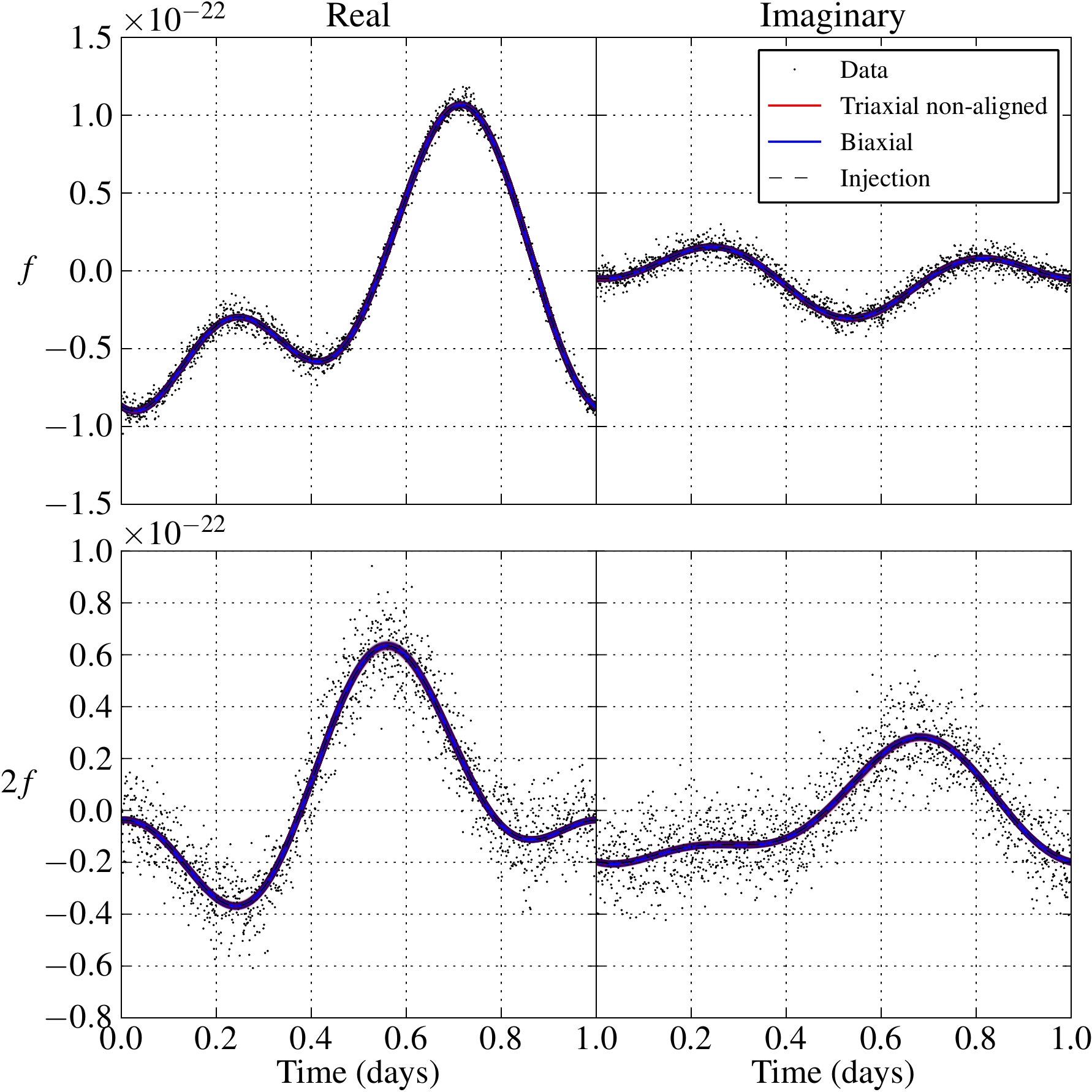}
 \end{center}
 \caption{\label{fig:waveform0} The real and imaginary waveforms for the $f$ and $2f$ data streams
of an \snr 500 signal for which the \biax model is favoured over the \tn model by $\roughly
\rme^{5.9}$.
The black points represent the simulated data, the overlapping red and blue lines show a
distribution of waveforms drawn randomly from the posterior parameter distributions for the \tn and
\biax models respectively, and the dashed black line shows the injected waveform. In this case the
lines are nearly impossible to distinguish on the plot, as a consistent waveform is recovered for both models.}
\end{figure}

We see that up to \snrs of $\roughly 10$ the simpler \ta model is more often favoured,
whilst at greater \snr the \tn and \biax models become more probable, i.e.\ the fact that the signal looks
more like the \tn and \biax models starts to overcome the Occam factors disfavouring them. For
signals with coherent \snrs of $\roughly 20$ we see that the simpler \ta model is still favoured
$\roughly 30\%$ of the time. This is mainly a result of the distribution of our population of simulated
signals for which, when the coherent \snr $\roughly 20$, about a third of the signals have an \snr
of $\lesssim 5$ in the $f$ component. Therefore, in these cases the $f$ component is providing very
little additional evidence for the \tn, or \biax models, and the simpler \ta model is being
favoured. A striking point to note is that out to \snrs of somewhere between 20 and 50 the \biax
model is favoured more than 50\% of the time over the true \tn model.  As noted in
Appendix~\ref{sect:relating_biaxial}, the \biax model is just a special case of the \tn model with
the constraint that $\Phi^C_{22} = 2\Phi^C_{21}$, so in some small fraction of the simulations
 this criterion will be fulfilled and the Occam factor will result
in the \biax model being favoured (see discussion below and  Fig.~\ref{fig:waveform0}). However, this does not account for the fraction of the cases that
the \biax model is favoured. To explain that we must look at another degeneracy: when the signal is
circularly polarised $\Phi^C_{22}$ and $\psi$ become very highly correlated (see, e.g.,
Fig.~\ref{fig:pdfsignalwaveformcirc}), and in these cases a combination of $\Phi^C_{22}$ and
$\psi$ can be found such that $\Phi^C_{22} \approx 2\Phi^C_{21}$. Therefore these will again look
like \biax signals and the Occam factor will start to favour them. This is demonstrated in
Fig.~\ref{fig:cosiota}, which shows the parameter posterior probability distributions for a \tn
signal with \snr $\roughly 20$ and $|\cos{\iota}| \roughly 0.5$, even with $|\cos{\iota}|$ some way from
unity the correlation between $\Phi^C_{22}$ and $\psi$ is still strong. This means that for \snrs
$\roughly 20$ about half the population will still be able to support the \biax model. Indeed in the
case of the signal in Fig.~\ref{fig:cosiota} the \biax model is {\it very} slightly favoured over
the
\tn model by a factor of $\roughly 1.3$. Even at \snrs as high as 500 this effect still means that
$\roughly
15\mbox{--}20\%$ of \tn signal simulations favour the \biax model.

\begin{figure}
 \begin{center}
  \includegraphics[width=0.47\textwidth]{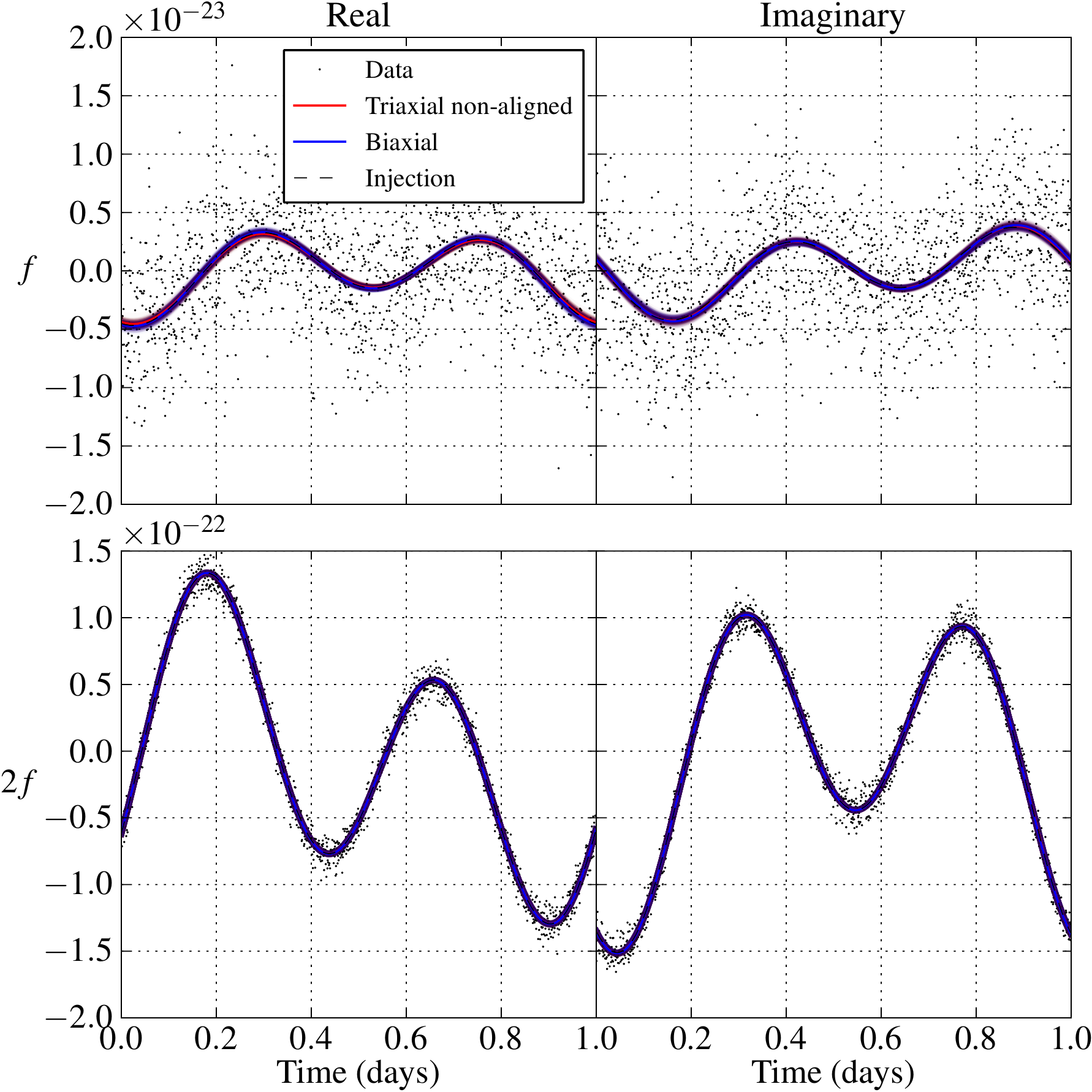}
 \end{center}
 \caption{\label{fig:waveform2} The real and imaginary waveforms for the $f$ and $2f$ data streams
of an \snr 500 signal for which the \biax model and \tn models are equally likely. The black points
represent the simulated data, the overlapping red and blue lines show a distribution of waveforms
drawn randomly from the posterior parameter distributions for the \tn and \biax models respectively,
and the dashed black line shows the injected waveform. In this case the lines are hard to
distinguish between as a consistent waveform is recovered for both models, although in the $f$ data
stream there is a minor discrepancy.}
\end{figure}

\begin{figure}
 \begin{center}
  \includegraphics[width=0.47\textwidth]{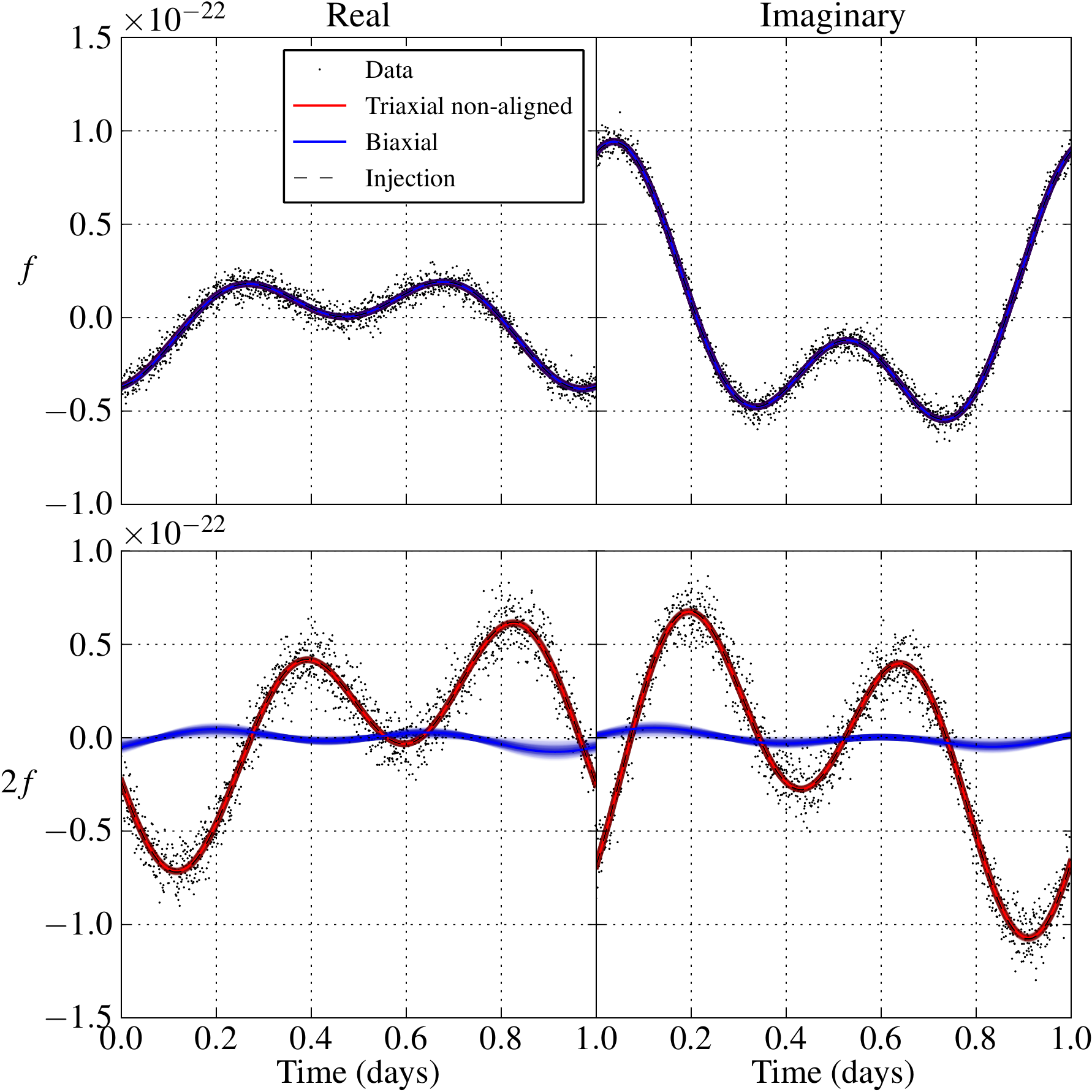}
 \end{center}
 \caption{\label{fig:waveform1} The real and imaginary waveforms for the $f$ and $2f$ data streams
of an \snr 500 signal for which the \biax model is disfavoured over the \tn model by a factor of
$\roughly \rme^{4690}$. The black points represent the simulated data, the overlapping red and blue
lines
show a distribution of waveforms drawn randomly from the posterior parameter distributions for the
\tn and \biax models respectively, and the dashed black line shows the injected waveform. In this
case the \biax model has to sacrifice any close fit in the $2f$ data-stream over producing a
very good fit to the higher \snr $f$ data-stream.}
\end{figure}

We can also see this effect by looking at three illustrative waveforms, each corresponding to
signals with \snr of 500, in the first case picking out a waveform where
 the \biax model is favoured by $\roughly \rme^{5.9}$,  in the second case the  \biax and \tn
models are equally likely, and in the the third case  the \biax model is strongly disfavoured by a
factor
of $\roughly \rme^{4690}$. The first case is shown in Fig.~\ref{fig:waveform0}, which happens to be a
simulation in which $\Phi^C_{22} \approx 2\Phi^C_{21}$ and $\cos{\iota} = 0.23$, so the waveform is
essentially a \biax waveform and both the \biax and \tn models recover the correct waveform almost
perfectly -- the parameters are also recovered consistently for both models. In this case the \tn
model provides no extra information, so the Occam factor means that the \biax model is favoured. The second case
is shown in Fig.~\ref{fig:waveform2}, in which $\Phi^C_{22} \not\approx 2\Phi^C_{21}$, but
$\cos{\iota} = -0.93$. In this case almost identical waveforms are recovered for both models, but
due to the $\Phi^C_{22}$ and $\psi$ degeneracy the recovered best fit parameters are not the same.
The third case is shown in Fig.~\ref{fig:waveform1}, where again $\Phi^C_{22} \not\approx
2\Phi^C_{21}$, but $\cos{\iota} = 0.23$. This shows that whereas the \tn model produces an excellent
fit in both the $f$ and $2f$ data-streams the \biax model sacrifices any attempt at a good fit in
the $2f$ data in favour of getting a very good fit in the higher \snr $f$ data.

\subsubsection{Detecting \ta signals}

Using the \ta injections we have again calculated odds ratios for each model. The efficiencies of
detecting these signals for each of the models, using the 1\% false alarm probability thresholds
given in Section~\ref{sec:noiseonly}, are shown in Fig.~\ref{subfig:efftri}.

The \ta model is just a special case of the two other models and we see that, as it is the simplest
model,
the best efficiency is achieved when we just use that model. However, the efficiency increase is
relatively small compared to more complex models, with those models giving almost
100\% efficiency for \snr greater than $\roughly 7$.

The main advantage of just performing the analysis assuming the \ta signal model is speed.
Using the waveform parameterisation the analysis in this case is $\roughly 2$ times
faster
than using the \tn model (the speed ratio does not vary significantly with \snr as the likelihoods
in both cases are fairly simple). However, if the signal does contain
significant power
at $f$ then, as we have seen in Figs~\ref{subfig:efffull} and \ref{subfig:triaxial}, assuming a
purely \ta signal model could lead to signals being missed.

The signal model comparison is shown in Fig.~\ref{subfig:modcomptri}. This shows that when a
purely \ta signal is present then that model is always favoured over the more complex models. As we
have noted earlier the \ta model is just a special case of the other models, so this result is
purely
down to the Occam factor rather than it being a better fit to the data. We also see the Occam
factor in play when comparing the \biax and \tn
models,
with the \biax model being favoured for the majority of signals. There is a slow trend towards the
\tn
model being more favoured. The reasons for this are not entirely clear, but a possible explanation is that  the
extra free parameter in the \tn model will allow it to more easily accommodate the necessary lack of
signal in the $f$ data-stream.

\subsubsection{Detecting \biax signals}

Finally, using the \biax simulations we have calculated odds ratios for each model. Fig.~\ref{subfig:effbi}
shows the efficiencies
of detecting these signals for each of the models, using the 1\% false alarm probability thresholds
given in Section~\ref{sec:noiseonly}. The results are
similar to those for the \tn injections of Fig.~\ref{subfig:efffull}, with the \ta model being
somewhat less efficient than the \biax or \tn models, due to its inability to detect
signals with low \snr in just the $2f$ component.  Again we see that the efficiencies of the \tn and
\biax models are very similar, despite one of them (the latter, in this case) being the correct
model.

The signal model comparison is shown in Fig.~\ref{subfig:modcompbi}.  In common with the cases
described above, at low \snr the \ta model is favoured over the other two models.  As the \snrs
increase, the \biax and \tn models become more favoured over the \ta model, by approximately
equal amounts, and are more often preferred for \snrs above about $10$. The \biax model is favoured
over the \tn model by a constant amount of $\roughly 90\%$ over all \snrs except the very smallest ones.

\section{Search in real data}\label{sec:S5results}

We performed a search for \gws from a selection of isolated (i.e.\ non-binary system) known pulsars using data from
the fifth LIGO science run (S5) \citep{Abbott:2007kv}. This is the first \gw search targeted at known pulsars to
include an explicit search for a component at the rotation frequency.  We  consider  43 pulsars that had
previously been targeted using an analysis only sensitive to the \ta model with emission at
twice the pulsars' rotation frequencies \citep{Abbott:2010}. We used science mode data from the two
LIGO Hanford Observatory detectors (H1 and H2) and the LIGO Livingston Observatory detector (L1)
covering 4 November 2005 to 1 October 2007, which was processed into sets of discrete Fourier
transforms using 30-min sections of data \citep[called short Fourier transforms, or SFTs; as used in e.g.][]{2013PhRvD..87d2001A}.
This gave a total of $\roughly 491$ days of data for H1, $\roughly 497$ days for H2 and $\roughly
392$ days for
L1. For each pulsar, we filtered the SFTs from each detector using a ``spectral interpolation''
routine \citep{Davies:2015} to create two narrow-band complex data streams sampled at one per 30
minutes: one with the phase evolution at the pulsars' rotation frequency removed, and the other with
the phase evolution at twice the pulsars' rotation frequency removed. For each data stream we also produced
estimates of the time-varying standard deviation of the corresponding noise.

For each pulsar we performed parameter estimation and calculated the evidence for each of the
\ta, \biax and \tn models. This allows us to perform model comparison for each source, assess the
detection of signals and produce 95\% credible region upper limits (bounded at zero) on the waveform model amplitudes $C_{21}$
and $C_{22}$. We show the results, based on a joint analysis of data from all three detectors, in Table~\ref{tab:S5results}.
As with the analysis in \citet{Abbott:2010} we expect amplitude
calibration uncertainties of 10\% for H1 and H2 and 13\% for L1. The table includes results for
the \ta model giving upper limits on the conventionally quoted \gw strain amplitude
$h_0$ (which in this model is related to $C_{22}$ via $h_0 = 2C_{22}$). These upper limits are
broadly consistent (within $\roughly 25\%$\footnote{A notable outlier is J1748$-$2446ac for
which
our new result is a factor of 1.7 times better than that in \citet{Abbott:2010}. This seems to be
due to there being a wandering spectral line feature in the H1 data close to $2f$ for this pulsar,
which the narrower bandwidth and noise estimation procedure of the spectral interpolation method is
able to veto, but that had artificially biased the noise level on the original result.}) with those in
\citet{Abbott:2010}, but are not identical due to differences in the processing pipeline (heterodyned,
averaged and low-pass filtered \citep{Dupuis:2005} vs.\ Fourier-transformed) and the fact that
here we have used a Gaussian likelihood for the data given our model (using a
noise estimate for each sample), rather than a Student's $t$ likelihood with its implied marginalisation over an
unknown noise level. From Table~\ref{tab:S5results} we see that the \ta model is favoured by
factors of $\roughly \rme^{12}$ in all cases.  Given that the previous gravitational wave search
reported in \citet{Abbott:2010} found no evidence for triaxial aligned signals (purely based on
examination of posterior probability distributions of the estimated signal amplitudes), we can
conclude that there is no evidence for \biax or \tn signals either. However, below we will also
assess the significance of the odds ratio for the \tn signal compared to noise, which appears in
the results table.

\begin{figure*}
 \begin{center}
  \includegraphics[width=1.0\textwidth]{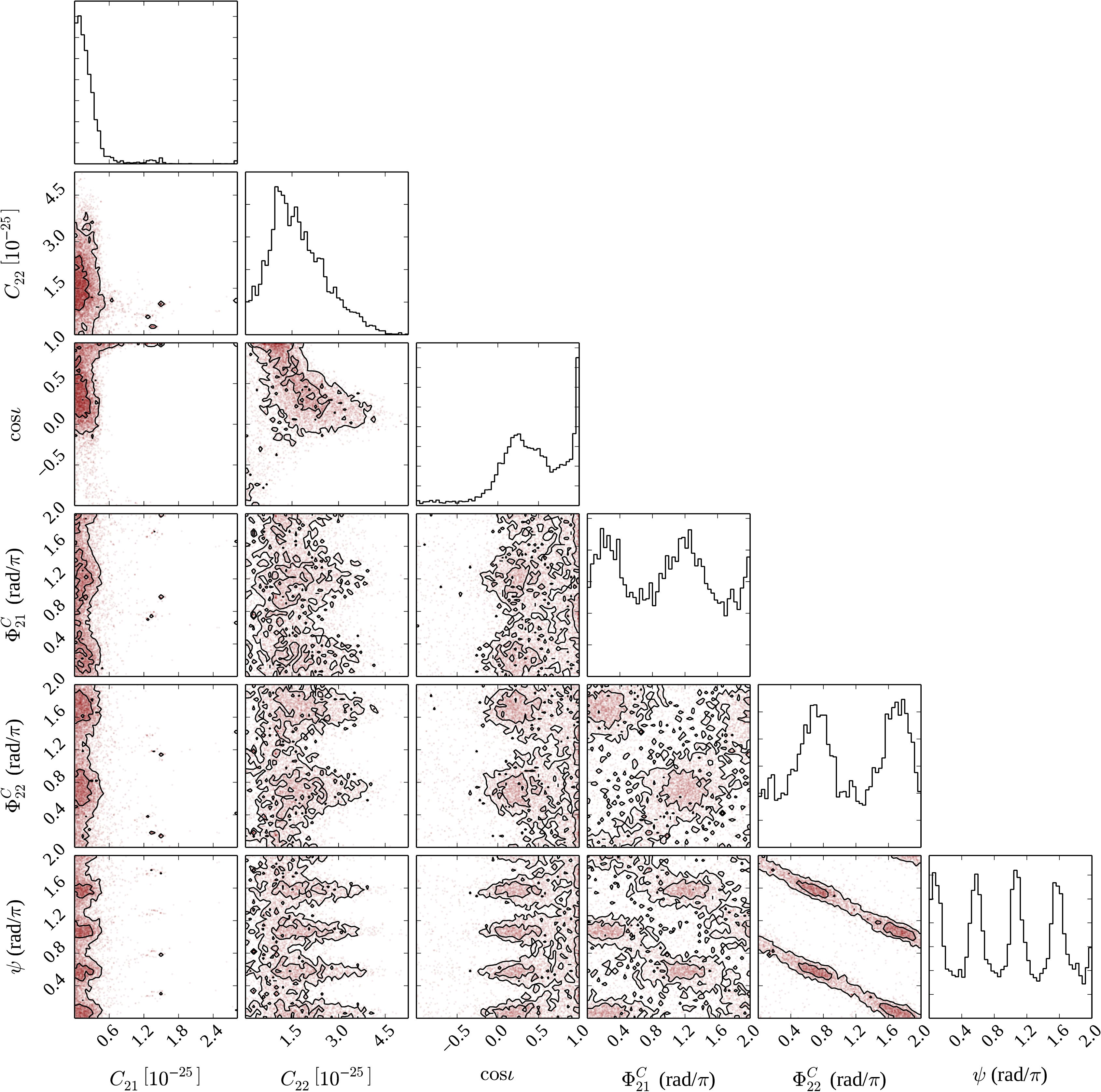}
 \end{center}
 \caption{\label{fig:S5example} The posterior probability distributions for the waveform
parameters of the \tn model found using S5 data for pulsar J0024$-$7204C.}
\end{figure*}

One advantage of using the waveform parameterisation over the source parameterisation is that
$C_{21}$ and $C_{22}$ directly represent the search sensitivity at $f$ and $2f$ respectively, whereas $I_{31}$ and
$I_{21}$ contribute to both harmonics in a complicated way that cannot be disentangled. However, it is
important to note that we currently do not have a good understanding of the physical interpretation
of $C_{21}$ and $C_{22}$. It is interesting to note from our results in Table~\ref{tab:S5results}
that, although in many cases the detector sensitivities are better at $f$ than $2f$
\citep[see e.g.\ Fig.~4 of][]{Abbott:2010}, we get smaller upper limits on $C_{22}$ for all bar two
pulsars (J0024$-$7204C and J2322$+$2057). This because when $\iota = 0$ the $f$ signal is
zero and $C_{21}$ can extend to arbitrarily large values, which creates a tail on the $C_{21}$
posterior probability distribution leading to the larger upper limits. The tail would be suppressed if a
Jeffreys prior were used for the amplitudes, but would still be present at some level as
the correlation is a feature of the waveform. Fig.~\ref{fig:S5example} shows an example of the posterior probability distributions
for the \tn model for pulsar J0024$-$7204C.

\subsection{Results significance}\label{sec:significance}

It is useful to have a way to assess whether the signal vs.\ noise odds ratio value for a particular pulsar is
large enough to be considered a detection (or detection candidate). As mentioned earlier, the odds
ratio itself tells us how much more probable the signal model is compared to noise given the data,
but fluctuations in this value for different noise realisations and the effect of our large
amplitude prior ranges mean that an understanding of the potential distribution of values is useful
in making a detection decision. We did this in Section~\ref{sec:noiseonly}
using simulations of noise-only data to get a distribution of odds ratios when no
signal was present, from which we could set a detection threshold, given a chosen false alarm rate.
For data in which the noise is purely Gaussian
this is straightforward, but with real data we need a way of producing the corresponding noise-only data with
the same statistics to get a representative distributions of odds ratios. It is also
useful to look at extra information such as the \snr of the maximum a posteriori recovered waveform.
Assessing a particular search's significance using an empirically estimated `background' distribution
of some detection statistic versus \snr is common for many searches for transient \gws, where in those
cases the `background' is generated through many time-slides of the data.

\begin{figure}
 \begin{center}
  \includegraphics[width=0.49\textwidth]{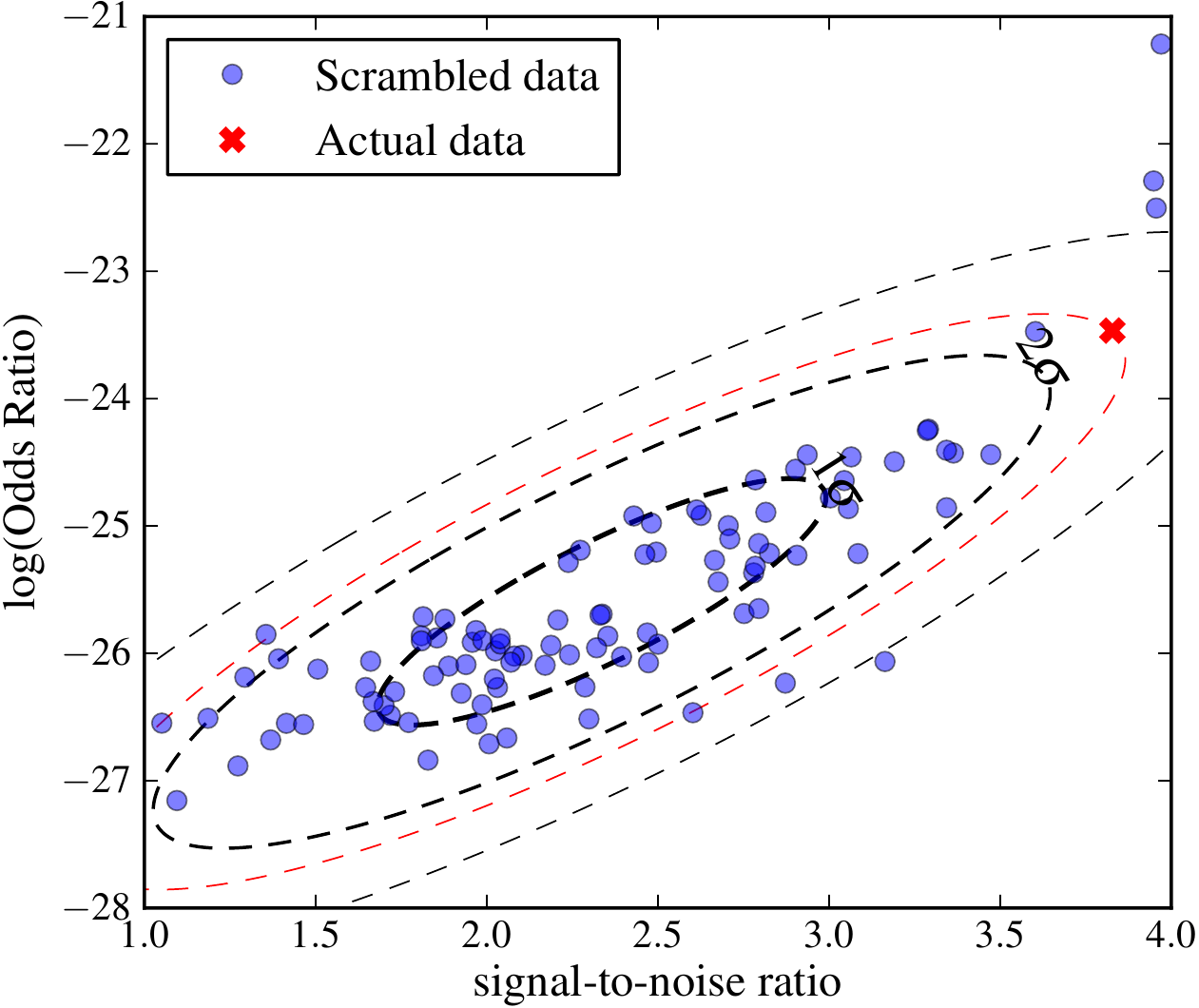}
 \end{center}
 \caption{\label{fig:background} The distribution of log odds ratios (for the \tn model versus
noise) against \snr for 100 `scrambled' noise-only realisations of the dataset for pulsar
J1024$-$0719. The
actual (i.e.\ not randomised in time) value is plotted as the red cross. Assuming that the
noise-only distribution is a bivariate Gaussian, the 1, 2 and $3\sigma$ (or $68.3\%$, $95.4\%$ and
$99.7\%$) probability contours are also plotted, along with the probability contour on which the
actual value lies.}
\end{figure}

To estimate a noise-only distribution of log odds ratios for the \tn model versus noise for each
pulsar we have made use of the same data as for the real analysis, but have `scrambled' the data by
randomly shuffling the
time order. This preserves the same noise statistics, but would completely de-cohere any signal
present, essentially giving us random realisations of the data. For each pulsar therefore, we
shuffled the data 100 times and calculated the log odds ratio for the \tn model versus Gaussian
noise, whilst also recording the \snr of the maximum a posteriori recovered signal model. We
calculated the correlation matrix of these 100 pairs of log odds ratios and \snr and then, assuming
their distribution is a bivariate Gaussian, calculated a set of probability contours in the odds
ratio--\snr plane for these.
The actual odds ratio and \snr for the pulsar (obtained using the non-scrambled data) can then be
placed on the plot, and its location relative to the noise-only distribution's contour lines gives an
indication of its significance.  For instance, an actual observation with a percentile contour
value $\gtrsim 99.7\%$
would be outside $3\sigma$ of the noise-only distribution (under the assumption of a bivariate Gaussian distribution). Of our
results the actual value lying on the highest percentile is for J1024$-$0719 at 98.05\% (or
equivalently at $2.34\sigma$ from the mean of the scrambled data distribution). This is shown in
Fig.~\ref{fig:background}.  Unsurprisingly, given the use of 100 scrambled data realisations there
are three points further out in the distribution, and we can conclude that this is not a
significant event (although we note that in this case
the distribution of scrambled data points is not quite a bivariate Gaussian as the three outlier points are beyond the
$3\sigma$ contour).

The use of the \snr in these plots provides some further level of discrimination from interference compared to
real signals in that data could return a high \snr signal (e.g.\ from a spectral line) that has a low odds ratio
due to not matching the signal model or being incoherent between detectors. Such a situation would give an obvious
outlier on plots such as Fig.~\ref{fig:background}. Real signals would be expected to have large values of  both odds ratio and \snr
and thus lie roughly along the diagonal of such plots.

\input resultstable.tex

\section{Conclusions} \label{sect:conclusions}

We have investigated detection and parameter estimation issues for the model of \gw emission from
rotating neutron stars proposed in \citet{Jones:2010}.   The model is based on the star having a
triaxial crust, coupled to an interior superfluid.  In the generic case of a \tn star, there is
emission at both the spin frequency $f$ and at $2f$.  We have also considered two special cases, of
a \biax star (also with emission at $f$ and $2f$), and a \ta one (emission only at $2f$), the last of these
being the case conventionally assumed in \gw searches.

We have found that in the generic case of  emission from a \tn star, the set of physical parameters
originally used in \citet{Jones:2010}, the `source parameters', are correlated in a highly complex
way. However, a re-parameterisation in terms of complex waveform amplitudes  using the `waveform
parameters' described in \citet{Jones:2015} breaks this degeneracy. When using a stochastic sampling
method (such as nested sampling) to estimate parameter probability distributions from data containing such a signal,
the complexity of the source parameter space makes a search there roughly half as computationally efficient as one in the waveform parameter space.

For a signal described by the \tn model, we showed that estimates of many of the true individual source parameters, including the important
parameters giving the asymmetry of the moment of inertia tensor, will always be poorly constrained
due to the degeneracies in the model. This may limit the astrophysical information that can be extracted on
such a source. We also note the (often overlooked) fact when discussing parameter estimation for these sources
that for any signal there is a degeneracy in the full physically allowed parameter space that means the signal
can only ever be constrained to a number of equally likely modes.

Working in the waveform parameterisation, and assuming stationary Gaussian noise in the data, we
have used simulations to calculate the odds ratio for three different signal models compared to
noise alone. We find that for a 1\% false alarm rate, calculated from an odds ratio threshold
value, all three models have efficiencies of close to 100\% for purely \ta signals with \snr
$\gtrsim 6$. The simplicity of the \ta model compared to the others does make it slightly more
efficient in this case, but only marginally so. For signals from the \tn and \biax models, with
significant power at $f$, when assuming either of those models, efficiencies are close to 100\% for
\snrs $\gtrsim 6$, while assuming the \ta model leads to some loss in efficiency, due to some
proportion of signals having very little power at $2f$. Without some idea of the true distribution
of signals strengths between $f$ and $2f$ we cannot say whether only searching a $2f$ would
cause any real signals to be missed. But, the ease of searching at both frequencies, and the only very minor
efficiency loss, makes performing such a search seem sensible in the future.

When comparing model evidences we find that for simulations containing any of the three models, at
very low \snr ($\lesssim 2.5$) the \ta model is always favoured. If the simulated signal is from
the \ta model then this is always favoured over the other two models, whilst there is a strong
preference for the  \biax as compared to the \tn one due to the \biax model being the simpler one.
For simulations containing \tn signals, the
\tn and \biax models are favoured over the \ta model for coherent \snrs $\gtrsim 10$. We see a similar
situation in simulations containing \biax injections: at low \snr the \ta model is
favoured, but at higher \snr the \biax and \tn models become favoured over the \ta one, while the
biaxial model becomes favoured over the \tn one.

Our results show that, even though to detect all \tn (or \biax) signals at \snr 20 one
should use the \tn (or \biax) model when computing the evidence, the Occam factor still
significantly penalises a reasonable percentage of those models when deciding which best fits the
data. As such it is worth noting that even at high \snr it is
often not possible to distinguish a \tn signal from a \biax one. However, the cost of searching
for a \tn signal compared to a \biax signal is relatively minor, so there is no reason to not
include such a search in the future.

Having developed the machinery needed to search for such signals, we then applied our methods to
real gravitational wave detector data.  Specifically, data from the S5 LIGO science run was used to
search for two-harmonic signals from 43 known pulsars with accurately known timing solutions, whose
spin frequencies lie within the LIGO band.  We found no gravitational wave signals, and so upper
limits were given on the amplitude-like parameters $C_{21}$ and $C_{22}$ on the triaxial non-aligned
model.  This is the first time gravitational wave detector data has been searched for such signals.
The techniques developed here will be applied to the more sensitive data soon to come from  the new
generation of advanced gravitational wave detectors \citep{2015CQGra..32g4001T,
2015CQGra..32b4001A}.

There is further work to do on the choice of prior probability distributions that one assumes for the parameters, particularly for the amplitude-like parameters.  A simple choice, uniform up to some fixed maximum amplitude, was used here, but other choices are possible and will affect the results obtained.  Closely related to this is the issue of the physical interpretation of the parameters
$C_{21}$ and $C_{22}$;  it is useful to consider whether they have a direct physical interpretation.  Their maximum values are
presumably related to the shear modulus and breaking strain of the crust, and also to the strength
of the interaction between the superfluid and non-superfluid parts of the star, but the relation is
not obvious, and is clearly worthy of further study.

\section*{Acknowledgements}

The simulations used in this paper were performed on the ARCCA cluster at Cardiff University, the resources for which were funded by an STFC grant (ST/I006285/1) supporting UK Involvement in the Operation of Advanced
LIGO. The data processing for
the results using real data were performed in the Atlas cluster at the Albert-Einstein-Institute in Hannover.
MP and GW acknowledge support from the STFC via grant number ST/L000946/1. DIJ acknowledges support from the STFC via grant
number ST/H002359/1, and also travel support from NewCompStar (a COST-funded Research Networking Programme).
We would like to thank the continuous waves working group of the LSC-Virgo Collaboration for useful discussions and
for preparation of the Fourier transformed data used in our analysis of real LIGO data. LIGO was
constructed
by the California Institute of Technology and Massachusetts Institute of Technology with funding from
the National Science Foundation and operates under cooperative agreement PHY-0757058.
This document has been assigned LIGO DCC number LIGO-P1400141.

\bibliography{multiharmonic}

\appendix
\section{Relating the source and waveform parameters}
\label{sect:relating_params}

In this appendix we briefly describe the relation between the source parameters and the waveform
parameters, for all three of our chosen physical models.  Full details can be found in
\citet{Jones:2015}.  Note that there a few small differences in notation between the equations given
here and those of \citet{Jones:2015}.  The angle $\psi$ of \citet{Jones:2015} is denoted by
$\lambda$ in this paper, as we reserve the symbol $\psi$ for the polarisation angle, describing the
projection of the star's spin axis on the plane of the sky.  Also, the quantities denoted by
$(\tilde C_{22}, \tilde C_{21})$ of \citet{Jones:2015} are denoted by $( C_{22},  C_{21})$ in this
paper.

\subsection{Triaxial non-aligned case} \label{sect:relating_tna}

For the triaxial non-aligned case, the waveform is given in terms of source
parameters by Eqns~(\ref{eq:1f}) and (\ref{eq:2f}), and by Eqns~(\ref{eq:1f_waveform})
and (\ref{eq:2f_waveform}) in terms of waveform
parameters.  It can be shown that the relationship between the five source
parameters $(I_{21}, I_{31}, \theta, \phi_0, \lambda)$ and the four source
parameters $(C_{21}, \Phi^{\rm C}_{21}, C_{22}, \Phi^{\rm C}_{22})$ is as
follows:
\begin{align}
\label{eq:C_22_source_params}
C_{22} =&   2
\{[I_{21}(\sin^2\lambda-\cos^2\lambda\cos^2\theta)- I_{31}\sin^2\theta]^2 + \nonumber \\
 & (I_{21}\sin2\lambda\cos\theta)^2\}^{1/2} ,\\
\label{eq:C_21_source_params}
C_{21} =&   2
\{(I_{21}\sin2\lambda\sin\theta)^2 + (I_{21}\cos^2\lambda - I_{31})^2\sin^2
2\theta\}^{1/2} ,
\end{align}
\begin{align}
\label{eq:Phi_22_source_params}
\tan[2\phi_0-\Phi^C_{22}] =& \frac{I_{21}\sin2\lambda\cos\theta}{I_{21}
(\sin^2\lambda-\cos^2\lambda\cos^2\theta)- I_{31}\sin^2\theta} ,\\
\label{eq:Phi_21_source_params}
\tan[\phi_0-\Phi^C_{21}] =&
\frac{(I_{21}\cos^2\lambda - I_{31})\sin2\theta}{I_{21}\sin2\lambda\sin\theta} ;
\end{align}
see equations (62)--(65) of \cite{Jones:2015}.  Given the set of five source parameters one can calculate unique values for the
four waveform parameters.  However, for a set of four waveform parameters, the
corresponding solution for the five source parameters will have one degree of
freedom, generating the sorts of complex structure in source parameter space
seen in Figs \ref{fig:pdfsignalsourcelinear} and
\ref{fig:pdfsignalsourcecirc}.

\subsection{Biaxial case} \label{sect:relating_biaxial}

If we set $I_{21} = 0$ in Eqns~(\ref{eq:1f}) and (\ref{eq:2f})  we obtain
the biaxial signal in terms of source parameters:
\begin{align}
h_+^{2f} =& -2 I_{31} (1+\cos^2\iota)  \sin^2\theta \cos2(\Omega t + \phi_0) ,\\
\label{eq:h_cross_biaxial}
h_\times^{2f} =& -4 I_{31}   \cos\iota  \sin^2\theta \sin2(\Omega t + \phi_0) ,
\\
\label{eq:h_plus_biaxial}
h_+^f =&  -I_{31}  \sin\iota \cos\iota  \sin 2\theta \sin(\Omega t + \phi_0) ,\\
h_\times^f =&  I_{31} \sin\iota  \sin 2\theta \cos(\Omega t + \phi_0) ,
\end{align}
where we have separated out the `$+$' and `$\times$' polarisation components.

It can be shown that the corresponding waveform parameterisation can be written  as
Eqn.~(\ref{eq:1f_waveform}) and (\ref{eq:2f_waveform}), with the extra
condition
\begin{equation}
\Phi^C_{22} = 2\Phi^C_{21} ;
\end{equation}
see equation (90) of \cite{Jones:2015}.  The relation between the source and waveform parameters can be shown to be
\begin{align}
\theta =& \tan^{-1}\left(\frac{2 C_{22}}{ C_{21}}\right)  ,\\
I_{31}  =&  -\frac{1}{2}  C_{22}  \left[1+\left(\frac{ C_{21}}{2   C_{22}}\right)^2\right]   , \\
 \phi_0 =& \Phi^C_{21} - \frac{\pi}{2} ;
\end{align}
see equations (82), (88) and (89) of \cite{Jones:2015}.

\subsection{Triaxial aligned case} \label{sect:relating_ta}

If we set $\theta = 0$ in Eqns~(\ref{eq:1f}) and (\ref{eq:2f})  we obtain
the triaxial aligned signal in terms of source parameters.  There in no signal at
frequency $f$, leaving only:
\begin{align}
\label{eq:h_plus_triaxial_GW}
h^{2f}_+ =& -2 I_{21} (1+\cos^2 \iota) \cos 2[\Omega t + (\phi_0+\lambda)] , \\
\label{eq:h_cross_triaxial_GW}
h^{2f}_\times =& -4 I_{21} \cos \iota \sin  2[\Omega t + (\phi_0 +\lambda)] .
\end{align}
The angles $\phi_0$ and $\lambda$ are now degenerate, with only their sum
appearing in the waveform.  This sum can now be taken as replacing the separate
parameters, or else one can simply set one of them to a constant value  and
search over the other (e.g. set $\lambda = 0$ and search over $\phi_0$).

The corresponding waveform parameterisation has $C_{21} = 0$ so that we only
have the $2f$ signal given by Eqn.~(\ref{eq:2f_waveform}), with the source
and waveform parameters being related by
\begin{align}
I_{21}  =& \frac{1}{2}  C_{22} ,\\
2(\phi_0 + \lambda) =& \Phi^C_{22}  ;
\end{align}
see equations (103) and (104) of \cite{Jones:2015}.

\section{Algorithm to fill in the full source parameter space}
\label{sect:algorithm}

In this appendix we present (as pseudo-code) the algorithm used to map the minimal source parameters space, whose ranges were given in Table \ref{tab:priorssource}, to the full parameter space.
\begin{algorithm}
\caption{Transformations required to map from the minimal range in the source model parameter space
to the full space. The values with the superscript `min' represent the original samples
from the minimal parameter range.}
\label{alg:source}
\begin{algorithmic}[1]
\For{$i \gets 1 \textbf{ to } 4$}
  \State draw a random unique set consisting of one quarter of the samples, such that $\bmath{j}_i$
is a vector of those samples' indices for loop $i$
  \State $\psi(\bmath{j}_i) = \psi^{\rm min}(\bmath{j}_i) + (i-1)\pi/2$
  \State $\theta(\bmath{j}_i) = \theta^{\rm min}(\bmath{j}_i)$
  \State $\phi_{0}(\bmath{j}_i) = \phi^{\rm min}_{0}(\bmath{j}_i)$
  \State $\lambda(\bmath{j}_i) = \lambda^{\rm min}(\bmath{j}_i)$
  \State $I_{21}(\bmath{j}_i) = I^{\rm min}_{21}(\bmath{j}_i)$
  \State $I_{31}(\bmath{j}_i) = I^{\rm min}_{31}(\bmath{j}_i)$
  \For{$k \gets 1 \textbf{ to } i$}
    \State $\theta_{\rm new} =
\arctan{\left(\frac{\sqrt{1-(\sin{\theta(\bmath{j}_i)}\sin{\lambda(\bmath{j}_i)})^2}}{\sin{
\theta(\bmath{j}
_i)}\sin{\lambda(\bmath{j}_i)}} \right)}$
    \State \begin{varwidth}[t]{\linewidth}
      $\phi_{0}(\bmath{j}_i) = $\par
\hspace{-10pt}
$\arctan{\left(\frac{\cos{\phi_{0}(\bmath{j}_i)}\cos{\lambda(\bmath{j}_i)} -
\sin{\phi_{0}(\bmath{j}_i)}\cos{\theta(\bmath{j}_i)}\sin{\lambda(\bmath{j}_i)}}
  {-\cos{\phi_{0}(\bmath{j}_i)}\cos{\theta(\bmath{j}_i)}\sin{\lambda(\bmath{j}_i)} -
\sin{\phi_{0}(\bmath{j}_i)}\cos{\lambda(\bmath{j}_i)}} \right)}$\par
      \end{varwidth}
    \State $\lambda(\bmath{j}_i) =
\arctan{\left(\frac{-\cos{\theta(\bmath{j}_i)}}{\sin{\theta(\bmath{j}_i)}\cos{\lambda(\bmath{j}_i)}}
\right)}$
    \State $I_{21}(\bmath{j}_i) = I_{31}(\bmath{j}_i) - I_{21}(\bmath{j}_i)$
    \State $I_{31}(\bmath{j}_i) = I_{31}(\bmath{j}_i)$
    \State $\theta_{i} = \theta_{\rm new}$
  \EndFor
\EndFor
\State randomly select half the samples, such that $\bmath{k}_1$ is a vector of those sample indices
\State apply $\lambda(\bmath{k}_1) = \lambda(\bmath{k}_1) + \pi$
\State randomly select half the samples, such that $\bmath{k}_2$ is a vector of those sample indices
\State apply $\theta(\bmath{k}_2) = \pi - \theta(\bmath{k}_2)$, $\phi_{0}(\bmath{k}_2) =
\phi_{0}(\bmath{k}_2) +\pi$ and $\lambda(\bmath{k}_2) =-\lambda(\bmath{k}_2)$
\State $\phi_{0} = \phi_{0}\ (\textrm{mod}\ 2\pi)$
\State $\lambda = \lambda\ (\textrm{mod}\ 2\pi)$
\end{algorithmic}
\end{algorithm}

\end{document}

%% file: abstract.tex
We investigate a method to incorporate signal models that allow
an additional frequency harmonic in searches for gravitational waves from
spinning neutron stars. We assume emission
is given by the general \tn model of Jones, whose
waveform under certain conditions reduces to that of a biaxial precessing
star, or a simple rigidly rotating \ta star. The \tn and \biax
models can produce emission at both the star's rotation frequency ($f$) and $2f$,
whilst the latter only emits at $2f$. We have studied parameter estimation
for signal models using both a set of physical {\it source parameters}, and a set of
{\it waveform parameters} that remove a degeneracy.
 We have assessed the signal detection
efficiency, and used Bayesian model selection to investigate how well
we can distinguish between the three models. We found that for signal-to-noise
ratios (\snrs) $\gtrsim 6$ there is no significant loss in efficiency if performing a
search for a signal at $f$ and $2f$ when the source is only producing
emission at $2f$. However, for sources with emission at both $f$ and $2f$
signals could be missed by a search only at $2f$. We also find that for a \ta source,
the correct model is always favoured, but for a \tn source it can be hard to
distinguish between the \tn model and the \biax model, even at high \snr.
Finally, we apply the method to a selection of known pulsars using data from
the LIGO fifth science run. We give the first upper
limits on \gw amplitude at both $f$ and $2f$ and apply the model selection
criteria on real data.

%% file: resultstable.tex
\begin{table*}
\centering
\def\m{\hbox{$\phantom{-}$}} 
\caption{\label{tab:S5results} Upper limits on \gw amplitudes, and log odds ratios comparing the \ta model (1),
\biax model (2) and \tn model (3) signal models, for 43 isolated pulsars using LIGO S5 data.
Also given is the percentile probability contour in the odds ratio (of \tn signal verses noise)---\snr plane of the background distribution on which the actual data point  sits.}
\begin{tabular}{@{}llllllllll}
\hline

PSR & $f$ (Hz) & $2f$ (Hz) & $C_{21}^{95\%}$ & $C_{22}^{95\%}$ & $h_{0}^{95\%}$ & $\ln{\mathcal{O}_{12}}$ &
$\ln{\mathcal{O}_{13}}$ & $\ln{\mathcal{O}_{23}}$ & percentile \\

\hline

J0024$-$7204C & $173.7$ & $347.4$ & $\ensuremath{6.5\!\times\!10^{-26}}$ & $\ensuremath{3.4\!\times\!10^{-25}}$ & $\ensuremath{6.5\!\times\!10^{-25}}$ & $12.8$ & $12.8$ & $-0.8$ &  94.5 \\
J0024$-$7204D & $186.7$ & $373.3$ & $\ensuremath{5.8\!\times\!10^{-26}}$ & $\ensuremath{2.0\!\times\!10^{-26}}$ & $\ensuremath{4.7\!\times\!10^{-26}}$ & $13.4$ & $13.4$ & $\m0.7$ &  81.3 \\
J0024$-$7204F & $381.2$ & $762.3$ & $\ensuremath{1.1\!\times\!10^{-25}}$ & $\ensuremath{4.7\!\times\!10^{-26}}$ & $\ensuremath{8.2\!\times\!10^{-26}}$ & $11.1$ & $11.1$ & $\m0.7$ &  46.0 \\
J0024$-$7204G & $247.5$ & $495.0$ & $\ensuremath{1.5\!\times\!10^{-25}}$ & $\ensuremath{4.2\!\times\!10^{-26}}$ & $\ensuremath{8.5\!\times\!10^{-26}}$ & $12.1$ & $12.1$ & $-1.4$ &  59.8 \\
J0024$-$7204L & $230.1$ & $460.2$ & $\ensuremath{6.4\!\times\!10^{-26}}$ & $\ensuremath{2.8\!\times\!10^{-26}}$ & $\ensuremath{5.9\!\times\!10^{-26}}$ & $12.5$ & $12.5$ & $\m0.3$ &  44.7 \\
J0024$-$7204M & $272.0$ & $544.0$ & $\ensuremath{7.2\!\times\!10^{-26}}$ & $\ensuremath{3.6\!\times\!10^{-26}}$ & $\ensuremath{7.4\!\times\!10^{-26}}$ & $12.6$ & $12.6$ & $-0.5$ &  43.2 \\
J0024$-$7204N & $327.4$ & $654.9$ & $\ensuremath{1.5\!\times\!10^{-25}}$ & $\ensuremath{3.2\!\times\!10^{-26}}$ & $\ensuremath{7.3\!\times\!10^{-26}}$ & $12.3$ & $12.3$ & $-0.1$ &  58.1 \\
J0711$-$6830 & $182.1$ & $364.2$ & $\ensuremath{9.5\!\times\!10^{-26}}$ & $\ensuremath{2.0\!\times\!10^{-26}}$ & $\ensuremath{4.2\!\times\!10^{-26}}$ & $12.1$ & $12.1$ & $-0.2$ &  23.2 \\
J1024$-$0719 & $193.7$ & $387.4$ & $\ensuremath{1.2\!\times\!10^{-24}}$ & $\ensuremath{3.3\!\times\!10^{-26}}$ & $\ensuremath{6.3\!\times\!10^{-26}}$ & $11.0$ & $11.0$ & $-0.3$ &  98.0 \\
J1730$-$2304 & $123.1$ & $246.2$ & $\ensuremath{7.8\!\times\!10^{-26}}$ & $\ensuremath{3.2\!\times\!10^{-26}}$ & $\ensuremath{6.6\!\times\!10^{-26}}$ & $13.1$ & $13.1$ & $\m1.3$ &  95.1 \\
J1744$-$1134 & $245.4$ & $490.9$ & $\ensuremath{6.7\!\times\!10^{-26}}$ & $\ensuremath{4.3\!\times\!10^{-26}}$ & $\ensuremath{8.3\!\times\!10^{-26}}$ & $13.2$ & $13.2$ & $\m0.8$ &  63.8 \\
J1748$-$2446C & $118.5$ & $237.1$ & $\ensuremath{6.2\!\times\!10^{-26}}$ & $\ensuremath{2.1\!\times\!10^{-26}}$ & $\ensuremath{4.4\!\times\!10^{-26}}$ & $12.4$ & $12.4$ & $\m0.7$ &  43.8 \\
J1748$-$2446D & $212.1$ & $424.3$ & $\ensuremath{3.8\!\times\!10^{-25}}$ & $\ensuremath{3.4\!\times\!10^{-26}}$ & $\ensuremath{7.5\!\times\!10^{-26}}$ & $12.7$ & $12.7$ & $-0.3$ &  36.7 \\
J1748$-$2446F & $180.5$ & $361.0$ & $\ensuremath{7.7\!\times\!10^{-26}}$ & $\ensuremath{3.1\!\times\!10^{-26}}$ & $\ensuremath{6.4\!\times\!10^{-26}}$ & $12.2$ & $12.2$ & $\m0.0$ &  60.0 \\
J1748$-$2446H & $203.0$ & $406.0$ & $\ensuremath{7.0\!\times\!10^{-26}}$ & $\ensuremath{3.5\!\times\!10^{-26}}$ & $\ensuremath{6.9\!\times\!10^{-26}}$ & $12.9$ & $12.9$ & $-0.2$ &  61.1 \\
J1748$-$2446K & $336.7$ & $673.5$ & $\ensuremath{2.1\!\times\!10^{-25}}$ & $\ensuremath{3.2\!\times\!10^{-26}}$ & $\ensuremath{7.0\!\times\!10^{-26}}$ & $11.3$ & $11.3$ & $\m0.1$ &  61.6 \\
J1748$-$2446L & $445.5$ & $891.0$ & $\ensuremath{6.3\!\times\!10^{-25}}$ & $\ensuremath{6.9\!\times\!10^{-26}}$ & $\ensuremath{1.7\!\times\!10^{-25}}$ & $11.4$ & $11.4$ & $-0.3$ &  81.0 \\
J1748$-$2446R & $198.9$ & $397.7$ & $\ensuremath{1.1\!\times\!10^{-25}}$ & $\ensuremath{3.8\!\times\!10^{-26}}$ & $\ensuremath{8.7\!\times\!10^{-26}}$ & $12.7$ & $12.7$ & $\m2.2$ &  60.0 \\
J1748$-$2446S & $163.5$ & $327.0$ & $\ensuremath{5.5\!\times\!10^{-26}}$ & $\ensuremath{2.6\!\times\!10^{-26}}$ & $\ensuremath{5.2\!\times\!10^{-26}}$ & $12.6$ & $12.6$ & $-0.7$ &  31.5 \\
J1748$-$2446T & $141.1$ & $282.3$ & $\ensuremath{8.3\!\times\!10^{-26}}$ & $\ensuremath{2.7\!\times\!10^{-26}}$ & $\ensuremath{5.4\!\times\!10^{-26}}$ & $13.7$ & $13.7$ & $\m0.2$ &  54.9 \\
J1748$-$2446aa & $172.8$ & $345.5$ & $\ensuremath{1.1\!\times\!10^{-25}}$ & $\ensuremath{9.9\!\times\!10^{-26}}$ & $\ensuremath{2.1\!\times\!10^{-25}}$ & $12.7$ & $12.7$ & $\m0.1$ &  63.0 \\
J1748$-$2446ab & $195.3$ & $390.6$ & $\ensuremath{6.0\!\times\!10^{-26}}$ & $\ensuremath{2.6\!\times\!10^{-26}}$ & $\ensuremath{5.1\!\times\!10^{-26}}$ & $13.1$ & $13.1$ & $\m0.8$ &  63.1 \\
J1748$-$2446ac & $196.6$ & $393.2$ & $\ensuremath{7.9\!\times\!10^{-26}}$ & $\ensuremath{2.3\!\times\!10^{-26}}$ & $\ensuremath{4.2\!\times\!10^{-26}}$ & $12.8$ & $12.8$ & $-0.1$ &  52.9 \\
J1748$-$2446af & $302.6$ & $605.3$ & $\ensuremath{5.5\!\times\!10^{-25}}$ & $\ensuremath{5.4\!\times\!10^{-26}}$ & $\ensuremath{1.1\!\times\!10^{-25}}$ & $12.0$ & $12.0$ & $-3.1$ &  95.1 \\
J1748$-$2446ag & $224.8$ & $449.6$ & $\ensuremath{7.5\!\times\!10^{-26}}$ & $\ensuremath{5.3\!\times\!10^{-26}}$ & $\ensuremath{1.1\!\times\!10^{-25}}$ & $12.3$ & $12.3$ & $\m0.3$ &  90.9 \\
J1748$-$2446ah & $201.4$ & $402.8$ & $\ensuremath{7.1\!\times\!10^{-26}}$ & $\ensuremath{3.1\!\times\!10^{-26}}$ & $\ensuremath{6.2\!\times\!10^{-26}}$ & $12.8$ & $12.8$ & $-1.0$ &  74.2 \\
J1801$-$1417 & $275.9$ & $551.7$ & $\ensuremath{1.2\!\times\!10^{-25}}$ & $\ensuremath{2.9\!\times\!10^{-26}}$ & $\ensuremath{5.8\!\times\!10^{-26}}$ & $11.7$ & $11.7$ & $-0.0$ &  29.4 \\
J1803$-$30 & $140.8$ & $281.6$ & $\ensuremath{6.9\!\times\!10^{-26}}$ & $\ensuremath{3.3\!\times\!10^{-26}}$ & $\ensuremath{6.5\!\times\!10^{-26}}$ & $11.7$ & $11.7$ & $\m0.9$ &  84.5 \\
J1823$-$3021A & $183.8$ & $367.6$ & $\ensuremath{8.7\!\times\!10^{-26}}$ & $\ensuremath{2.2\!\times\!10^{-26}}$ & $\ensuremath{4.1\!\times\!10^{-26}}$ & $14.1$ & $14.1$ & $\m0.6$ &  82.0 \\
J1824$-$2452A & $327.4$ & $654.8$ & $\ensuremath{8.4\!\times\!10^{-26}}$ & $\ensuremath{3.9\!\times\!10^{-26}}$ & $\ensuremath{8.4\!\times\!10^{-26}}$ & $13.2$ & $13.2$ & $\m0.8$ &  91.2 \\
J1824$-$2452B & $152.7$ & $305.5$ & $\ensuremath{7.7\!\times\!10^{-26}}$ & $\ensuremath{2.5\!\times\!10^{-26}}$ & $\ensuremath{5.2\!\times\!10^{-26}}$ & $13.6$ & $13.6$ & $-0.1$ &  94.3 \\
J1824$-$2452E & $184.5$ & $369.1$ & $\ensuremath{5.9\!\times\!10^{-26}}$ & $\ensuremath{3.5\!\times\!10^{-26}}$ & $\ensuremath{6.7\!\times\!10^{-26}}$ & $12.2$ & $12.2$ & $-0.3$ &  47.4 \\
J1824$-$2452F & $408.0$ & $815.9$ & $\ensuremath{1.4\!\times\!10^{-24}}$ & $\ensuremath{4.6\!\times\!10^{-26}}$ & $\ensuremath{9.1\!\times\!10^{-26}}$ & $11.9$ & $11.9$ & $\m0.6$ &  34.5 \\
J1843$-$1113 & $541.8$ & $1083.6$ & $\ensuremath{1.8\!\times\!10^{-25}}$ & $\ensuremath{7.6\!\times\!10^{-26}}$ & $\ensuremath{1.5\!\times\!10^{-25}}$ & $12.5$ & $12.5$ & $-0.2$ &  97.0 \\
J1905+0400 & $264.2$ & $528.5$ & $\ensuremath{6.2\!\times\!10^{-26}}$ & $\ensuremath{3.8\!\times\!10^{-26}}$ & $\ensuremath{7.8\!\times\!10^{-26}}$ & $12.4$ & $12.4$ & $\m0.6$ &  47.8 \\
J1910$-$5959B & $119.6$ & $239.3$ & $\ensuremath{5.9\!\times\!10^{-26}}$ & $\ensuremath{1.9\!\times\!10^{-26}}$ & $\ensuremath{4.2\!\times\!10^{-26}}$ & $12.8$ & $12.8$ & $-0.8$ &  49.3 \\
J1910$-$5959C & $189.5$ & $379.0$ & $\ensuremath{6.0\!\times\!10^{-26}}$ & $\ensuremath{2.2\!\times\!10^{-26}}$ & $\ensuremath{4.6\!\times\!10^{-26}}$ & $12.1$ & $12.1$ & $-1.9$ &  70.9 \\
J1910$-$5959D & $110.7$ & $221.4$ & $\ensuremath{7.7\!\times\!10^{-26}}$ & $\ensuremath{1.4\!\times\!10^{-26}}$ & $\ensuremath{2.9\!\times\!10^{-26}}$ & $12.6$ & $12.6$ & $\m1.3$ &  50.6 \\
J1910$-$5959E & $218.7$ & $437.5$ & $\ensuremath{6.3\!\times\!10^{-26}}$ & $\ensuremath{2.1\!\times\!10^{-26}}$ & $\ensuremath{4.6\!\times\!10^{-26}}$ & $12.4$ & $12.4$ & $\m0.4$ &  32.7 \\
J1911+1347 & $216.2$ & $432.3$ & $\ensuremath{2.3\!\times\!10^{-24}}$ & $\ensuremath{2.5\!\times\!10^{-26}}$ & $\ensuremath{6.0\!\times\!10^{-26}}$ & $12.2$ & $12.2$ & $-0.1$ &  51.7 \\
J1939+2134 & $641.9$ & $1283.9$ & $\ensuremath{2.4\!\times\!10^{-25}}$ & $\ensuremath{7.5\!\times\!10^{-26}}$ & $\ensuremath{1.6\!\times\!10^{-25}}$ & $11.4$ & $11.4$ & $-0.1$ &  37.2 \\
J2124$-$3358 & $202.8$ & $405.6$ & $\ensuremath{8.3\!\times\!10^{-26}}$ & $\ensuremath{2.1\!\times\!10^{-26}}$ & $\ensuremath{4.7\!\times\!10^{-26}}$ & $12.5$ & $12.5$ & $-0.4$ &  57.0 \\
J2322+2057 & $208.0$ & $415.9$ & $\ensuremath{4.3\!\times\!10^{-26}}$ & $\ensuremath{4.4\!\times\!10^{-26}}$ & $\ensuremath{8.8\!\times\!10^{-26}}$ & $13.9$ & $13.9$ & $-0.2$ &  52.5 \\
\hline
\end{tabular}
\end{table*}